\documentclass[preprint]{elsarticle}
\usepackage{lineno,hyperref,pdflscape}
\modulolinenumbers[5]

\usepackage{multicol}
\usepackage{color}
\usepackage{xspace}
\usepackage{enumerate}
\usepackage{amsmath} 
\usepackage{ulem} 
\usepackage{placeins}
\usepackage{threeparttable}

\usepackage{tabularx}
\usepackage{mathtools} 
\usepackage{amsfonts}
\newcolumntype{M}{>{\centering\arraybackslash}m{1.95cm}}
\usepackage[export]{adjustbox}
\usepackage{float}
\usepackage{braket}
\usepackage{lipsum}
\usepackage{longtable}
\graphicspath{{figure/}}
\newcommand\T{\rule{0pt}{3ex}}       
\newcommand\B{\rule[-1.5ex]{0pt}{0pt}} 

\usepackage{booktabs}
\usepackage{blindtext}
\makeatletter
\newcommand{\colorcaption}[2][]{%
	\begingroup%
	\renewcommand{\@caption@fignum@sep}{ (Color online). }%
	\caption[#1]{#2}%
	\endgroup%
}
\makeatother

\usepackage{amssymb}
\usepackage{graphicx}
\usepackage{rotating}
\usepackage{tikz}
\newcommand{\gj}[9]{ \begin{Bmatrix}
		#1 & #2 & #3 \\
		#4 & #5 & #6 \\
		#7 & #8 & #9
\end{Bmatrix}}
\newcommand{\Gj}[6]{ \begin{Bmatrix}
		#1 & #2 & #3 \\
		#4 & #5 & #6 
\end{Bmatrix}}

\begin{document}
	
\begin{frontmatter}
 \title{Nuclear structure properties of Si and P isotopes with the microscopic effective interactions}
\author{Priyanka Choudhary}
\author{Praveen C. Srivastava\footnote{Corresponding author: praveen.srivastava@ph.iitr.ac.in}}
\address{Department of Physics, Indian Institute of Technology Roorkee, Roorkee 247667, India}
\date{\hfill \today}
\begin{abstract}
In the present work, we have reported comprehensive study of the $sd$-shell nuclei Si and P with neutron number varying from $N = 9$ to $N = 20$ using the microscopic effective valence shell interactions, namely, N3LO, JISP16 and DJ16A. These effective $sd$-shell interactions are developed using the \textit{ab initio} no-core shell model wave functions and the Okubo-Lee-Suzuki transformation method. For comparison, we have also performed shell model calculations with the empirical USDB interaction. Energy spectra and electromagnetic properties of these isotopic chains have been studied. Theoretically calculated shell model results are compared with the available experimental data, to check the predictive strength of these microscopic interactions. It is found that the binding energies of the ground states are  better reproduced with the DJ16A interaction as compared to other microscopic interactions  and a proton subshell closure at $Z=14$ is obtained in Si. Spin-tensor decomposition of two-body interaction is presented to understand the contributions from central, vector and tensor components into these interactions. Spectroscopic strengths of $^{23}$Al($d$,$n$)$^{24}$Si are examined for the newly performed experiment at NSCL.
\end{abstract}
		

\end{frontmatter}
	
\section{Introduction}
Over the last few years nuclear structure study of $p$- and lower $sd$-shell nuclei within the framework of \textit{ab initio} no-core shell model (NCSM) approach has achieved a great amount of success \cite{BARRETT2013131,PhysRevC502841,PhysRevC542986,PhysRevC573119,PRC1906,PRC62,2009,MVS2009,stetcu1,stetcu2,PhysRevC.102.044309,PhysRevC.107.014309,Nucl.Phys.A,JPG,10.1093/ptep/ptz073,Saxena_2020}. In the NCSM calculations, all the constituents of a nucleus are considered to be active particles. No concept of inert core as such is assumed in this approach. When the NCSM approach is implemented to study the properties of  heavier-mass nuclei, dimension of the Hamiltonian matrix becomes  too large to handle with the currently accessible computational power.  So, other \textit{ab initio} approaches  such as in-medium similarity renormalization group (IM-SRG) \cite{PhysRevLett.113.142501,PhysRevLett.106.222502}, coupled cluster (CC) theory \cite{PhysRevLett.113.142502,PhysRevC.94.011301} and symmetry-adapted no-core shell model (SA-NCSM)  \cite{Launey, Panu1, Jonathan,SA-NCSM} are being used to describe the structure of heavier nuclei. Electromagnetic properties, spectroscopic factor strengths and Gamow-Teller (GT) strengths of $sd$-shell nuclei have been calculated using former two of the aforementioned \textit{ab initio} approaches in Refs. \cite{PhysRevC.96.024316, Collectivity, PhysRevC.105.034333, PhysRevC.105.034332, SF, GT}. Another \textit{ab initio} approach, known as  the \textit{ab initio} shell model with  a core, has also been developed in Refs. \cite{PhysRevC.78.044302,PhysRevC.91.064301,PhysRevC.100.054329}.

In  the \textit{ab initio} shell model with core approach, effective two-body interaction is constructed by performing \textit{ab initio} NCSM calculation in N$_\mathrm{max}$ model space for a nuclei  followed by explicit projection onto the 0$\hbar\Omega$ space using Okubo-Lee-Suzuki (OLS) transformation method. Thus, the effective Hamiltonian is separated out into inert core, one, and two-body terms. In  \cite{PhysRevC.91.064301}, the J-matrix inverse scattering potential (JISP16) \cite{SHIROKOV200733} and chiral next-to-next-to-next-to-leading order (N3LO) interactions \cite{PhysRevC.68.041001} for $sd$-shell have been developed and it is demonstrated that low-lying energy spectra of $^{18}$F using the NCSM is exactly reproduced with these derived effective interactions. The same interaction has been applied on $^{19}$F and  the shell model results  obtained therefrom are found to be very close to the NCSM results. It  has also been reported that the effective Hamiltonian has weak A dependence,  which means  that this interaction can be  applied to the heavier $sd$-shell nuclei. Further, new effective interactions, Daejeon 16 (DJ16) \cite{A.M.Shirokov} and its monopole corrected version Daejeon 16A (DJ16A),  have been constructed for $sd$-shell \cite{PhysRevC.100.054329} similar to the above mentioned approach. Smirnova \textit{et al.} \cite{PhysRevC.100.054329} have implemented these microscopic interactions on O isotopes, odd-A F isotopes, $^{26}$F, $^{22}$Na, $^{28,29}$Si, $^{32}$S, $^{39}$K and deformed rotor $^{24}$Mg. A remarkable consistency between theoretical description and experimental data is obtained \cite{PhysRevC.100.054329}. Success of these effective microscopic interactions  motivates us to fruitfully apply these interactions to study heavier nuclei in $sd$-shell. We extend the earlier work \cite{PhysRevC.100.054329} to study  Si and P isotopic chains in $sd$-shell region. Apart from energy spectra, we have also calculated the nuclear observables for these nuclei. Our present comprehensive study will add more information to the earlier work \cite{PhysRevC.100.054329}.

Nuclear structure properties of $sd$-shell nuclei including proton and neutron drip lines have been studied in several theoretical and experimental works \cite{Collectivity,P.D.Cottle,SPEs,Kaneko,Kumar:2021rsv,PhysRevC.74.034315,PhysRevC.78.064302,PhysRevC.101.064312}. In  \cite{Kumar:2021rsv}, authors have implemented relativistic Hartree Bogoliubov model with density dependent meson-exchange and pairing interactions to study nuclear shapes of Mg, Si, S and Ar. They found existence of magic numbers at $N = 8$ and $20$ with spherical shape in these isotopic chains and disappearance of $N = 28$ shell closure with a finite deformation.  Prolate-oblate shape coexistence in some isotopes of Mg, Si, S and Ar isotopic chains have also been observed. In  \cite{Collectivity}, it has been reported that excitation of nucleons from $sd$- to $pf$-shell plays an important role for ``island of inversion'' nuclei at $N = 20$. \textit{Ab initio} results with IMSRG and CC methods were seen to be in  reasonable agreement with the experimental data for the $sd$-shell nuclei Ne, Mg and Si except for $N = 20$.  Considerable success of the USDA and USDB interactions \cite{PhysRevC.74.034315,PhysRevC.78.064302} in describing the nuclear structure properties of $sd$-shell nuclei has led to the construction of new isospin-breaking USD-type interactions \textit{viz.} the USDC and USDI interactions \cite{PhysRevC.101.064312}. These new interactions improve predictions for separation energy in the entire $sd$-shell.

In  \cite{PhysRevC.95.021304}, a new type of \textit{ab initio} nucleon-nucleon (NN) interaction was developed for $sd - pf$ model space,  \textit{viz.} the EEdf1 interaction. 
This interaction was derived from the fundamental chiral effective field theory ($\chi$EFT) \cite{RMP,EFT1,EFT2,EFT3} based on Quantum chromodynamics (QCD) \cite{QCD} and extended Kuo-Krenciglowa (EKK) method \cite{TAKAYANAGI201191,TAKAYANAGI201161,PhysRevC.89.024313}. The Fujita-Miyazawa three nucleon forces \cite{10.1143/PTP.17.360} were also included. 
It was initially implemented on medium mass nuclei Ne, Mg and Si isotopes \cite{PhysRevC.95.021304}. Exotic neutron rich Ne, Mg and Si isotopes were explained in the aforementioned study by particle-hole excitations across two major shells without fitted interaction. Further, this interaction  has also been applied to determine the properties of exotic nuclei with $Z=9-12$ and $N$ up to driplines \cite{Tsunoda:2020gpt}. This work was  extended to determine the spectroscopic properties such as magnetic dipole and electric quadrupole moments, charge and matter radii for Na and Mg isotopes, in  Ref. \cite{PhysRevC.105.014319}. 

Spectroscopic factor strength describes the nature and occupancy of the single particle orbits in a nucleus that are used to determine the structure of nucleus. In Ref. \cite{PhysRevC.86.015806}, experimental spectroscopic factor strength for ground state (g.s.) of $^{24}$Si from $^{23}$Al($p,\gamma$)$^{24}$Si reaction was obtained. Recently, an experiment has been performed with a transfer reaction $^{23}$Al($d,n$)$^{24}$Si  at National Superconducting Cyclotron Laboratory \cite{PhysRevLett.122.232701} to measure excited states and their spectroscopic factor strengths that might be of astrophysical interest. Motivated with these recent experimental data, we have calculated theoretical spectroscopic factor strengths for g.s. as well as excited states of $^{24}$Si for one proton capture reaction $^{23}$Al($d,n$)$^{24}$Si using microscopic N3LO, JISP16 and DJ16A interactions, and have also carried out a comparison with the USDB results, in the present study. 

The present work is organized as follows.  In Section II, we briefly describe the methodology to develop the microscopic effective $sd$-shell interaction and details about the interactions used in our calculations. Next, we show the results of energy spectra of Si and P isotopic chains in Section III. Effective proton single-particle energies of Si isotopes have been carried out in Section IV. In Section V, the spin-tensor decomposition of the effective interactions and the monopole components of decomposed parts have been presented. We then discuss the electromagnetic properties of Si and P nuclei in Sections VI. Spectroscopic factor strengths for $^{24}$Si are studied in Section VII.  
Finally, we  draw conclusion of the present study in Section VIII. 

\section{Microscopic effective $sd$-shell interactions}
To describe the nuclear structure properties of $sd$-shell nuclei, we have performed the nuclear shell model calculations with microscopic effective $sd$-shell interactions in the present work. The detailed procedure to develop these effective $sd$-shell interactions has been described in Refs. \cite{PhysRevC.91.064301,PhysRevC.100.054329}, which we have briefly presented here.

In this approach, the microscopic effective valence shell interactions are derived from the \textit{ab initio} NCSM wave functions. The NCSM Hamiltonian for a nucleus of $A$ point like nucleons interacting through realistic interaction can be written as
\begin{equation}\label{(2)}
	H_A = \frac{1}{A}\sum_{i<j}^{A}\frac{(\vec{p}_i-\vec{p}_j)^2}{2m} + \sum_{i<j}^{A}V^{\text{NN}}_{ij},
\end{equation}
where, m is the mass of a nucleon. This Hamiltonian, which is translationally invariant, has relative kinetic energy term and two-nucleon interaction term including Coulomb interaction between protons. In these calculations three-body forces are omitted.

The harmonic oscillator (HO) basis states (Slater determinate basis) up to $N_\mathrm{max}$, which is maximum number of HO quanta above the unperturbed $A$ nucleon configuration, are used to solve the eigen value problem of $A$ nucleon system. Several realistic NN interactions, \textit{i.e.} charge-dependent Bonn 2000 (CDB2K) \cite{PhysRevC.63.024001} and chiral effective field theory interactions \cite{PhysRevC.68.041001} generate strong short-range correlations. To obtain convergent results, large basis states are required which is constrained by computational limitations. Thus, a renormalization method is employed to soften the standard realistic interactions. There are two renormalization techniques: OLS similarity transformation \cite{Prog.Theor.Phys.12,Prog.Theor.Phys.,Prog.Theor.Phys.68,Prog.Theor.Phys.92} and SRG method \cite{PhysRevC.75.061001}, with the former being implemented in the procedure outlined here. 

To make derivation easier, a frequency dependency is introduced in the NCSM calculations. Addition of center-of-mass HO Hamiltonian to the initial Hamiltonian \ref{(2)} modifies the resultant Hamiltonian as
\begin{equation}\label{(3)}
		H_{a} + H_{\text{c.m.}}  
		= \sum_{i=1}^{a} \bigg[ \frac{\vec{p}_i^{\,\,2}}{2m}+ \frac{1}{2}m{\Omega}^2 \vec{r}_i^{\,\,2} \bigg]+ \sum_{i<j=1}^{a} \bigg[ V_{ij}^{\text{NN}} - \frac{m {\Omega}^2}{2A} {(\vec{r}_i - \vec{r}_j)}^2 \bigg] . 
\end{equation} 
This $H_{\text{c.m.}}$ is subtracted out from the final calculations. If $a$ = $A$ then $H_{a}$ becomes the initial Hamiltonian \ref{(2)}. The effective NN interaction for the NCSM calculations is derived from Eq. \ref{(3)} in $a$ cluster approximation. In these NCSM calculations, the $a$ = 2 cluster approximation has been used. First OLS transformation is applied to construct a primary effective Hamiltonian for $A = 18$ system  with NCSM parameters $N_\mathrm{max}$ = 4 and $\hbar$$\Omega$ = 14 MeV.  The lowest 28 eigen states are calculated to perform second OLS transformation. This 18-body Hamiltonian is projected onto 0$\hbar$$\Omega$ model space. These eigen states of $^{18}$F are dominated by the configurations with $^{16}$O system in the lowest possible HO orbits and two nucleons in $sd$-shell. The secondary effective Hamiltonian reproduces exactly same energy of the states in $^{18}$F, as generated by the primary effective Hamiltonian. Further, the NCSM calculations have been performed for $^{16}$O, $^{17}$O and $^{17}$F with primary effective Hamiltonian to calculate core and one-body single-particle energies. These core and one-body components have been subtracted out from secondary effective Hamiltonian of $^{18}$F to obtain residual two-body matrix elements (TBMEs) for the $sd$ valence shell.

In this work, three microscopic  $sd$-valence space interactions have been used, \textit{viz.} N3LO, JISP16 and DJ16A. Together with these microscopic interactions, we have also used empirical USDB interaction \cite{PhysRevC.78.064302} to know how much reliable description of nuclei are obtained from the microscopic potentials. The calculated results with these interactions are compared with the experimental data. In the case of the empirical USDB interaction, single-particle energies are $\epsilon$(d$_{5/2}$) = -3.9257 MeV, $\epsilon$(s$_{1/2}$) = -3.2079 MeV and $\epsilon$(d$_{3/2}$) = 2.1117 MeV. As reported in Ref. \cite{PhysRevC.91.064301,PhysRevC.100.054329}, one-body terms derived in an \textit{ab initio} way for microscopic interaction are different from the phenomenological energies; the ordering of the orbitals $d_{5/2}$ and $s_{1/2}$ is reversed and the energy separation between orbitals $d_{5/2}$ and $d_{3/2}$ is larger than the USDB difference. Also, the single particle energies derived from \textit{ab initio} approach give rise the deficiencies in the nuclear spectra, mentioned in Ref. \cite{PhysRevC.100.054329}. Due to these issues, USDB single particle energies are taken into account for all these microscopic interaction in order to focus on the TBMEs. We have calculated only positive parity states of these nuclei. We have  performed nuclear shell model calculations using the code KSHELL \cite{KSHELL2019}.

Experimental g.s. energies of  $sd$-shell nucleus relative to the g.s. energy of $^{16}$O with the Coulomb energy  correction term is determined as:
\begin{equation}
	E(A,Z)^r = E(A,Z) - E(^{16}{\rm O})-E_c(Z),
\end{equation}
where, $E(A,Z)^r$ and $E(A,Z)$ are relative and absolute g.s. energy. $E(^{16}{\rm O})$ is the g.s. energy of $^{16}$O, which have the value of -127.619 MeV.  $E_c(Z)$ is the Coulomb correction energy  taken from Ref. \cite{PhysRevC.74.034315}.

\section{Results and discussion}

\begin{figure*}
	\includegraphics[width=7cm,height=5.5cm]{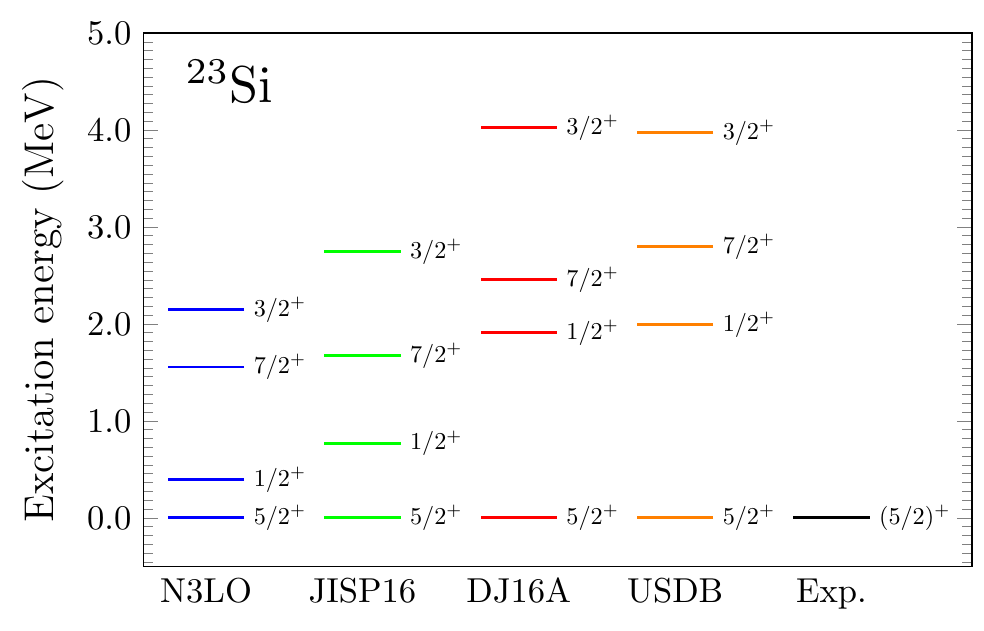}
	\includegraphics[width=7cm,height=5.5cm]{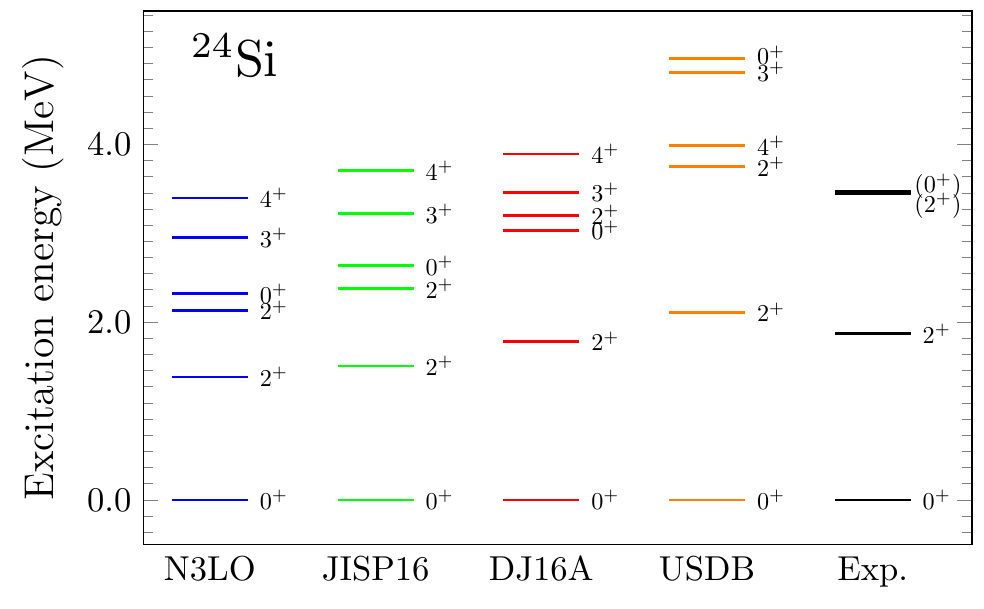}
	\includegraphics[width=7cm,height=5.5cm]{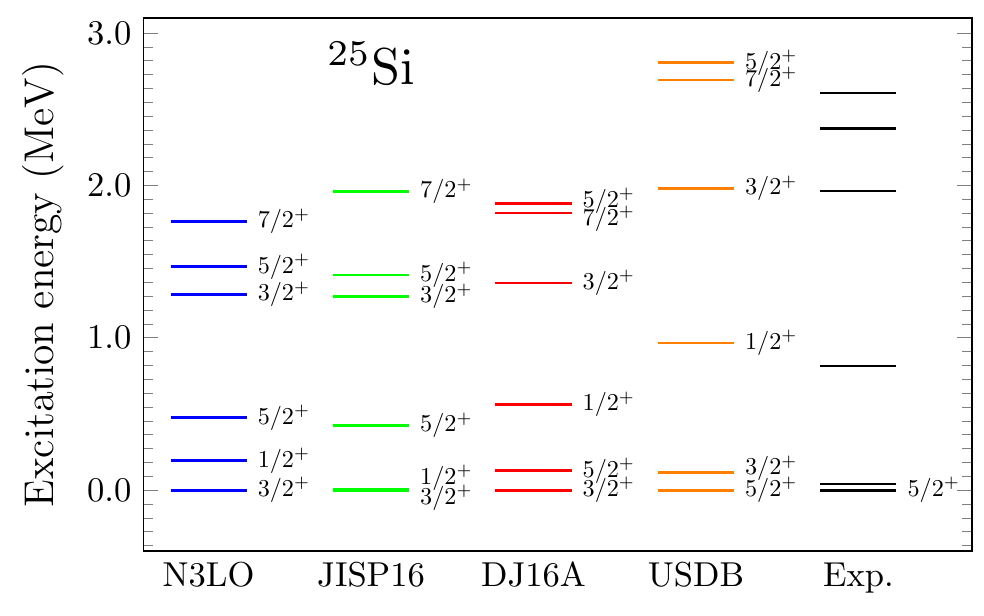}
	\includegraphics[width=7cm,height=5.5cm]{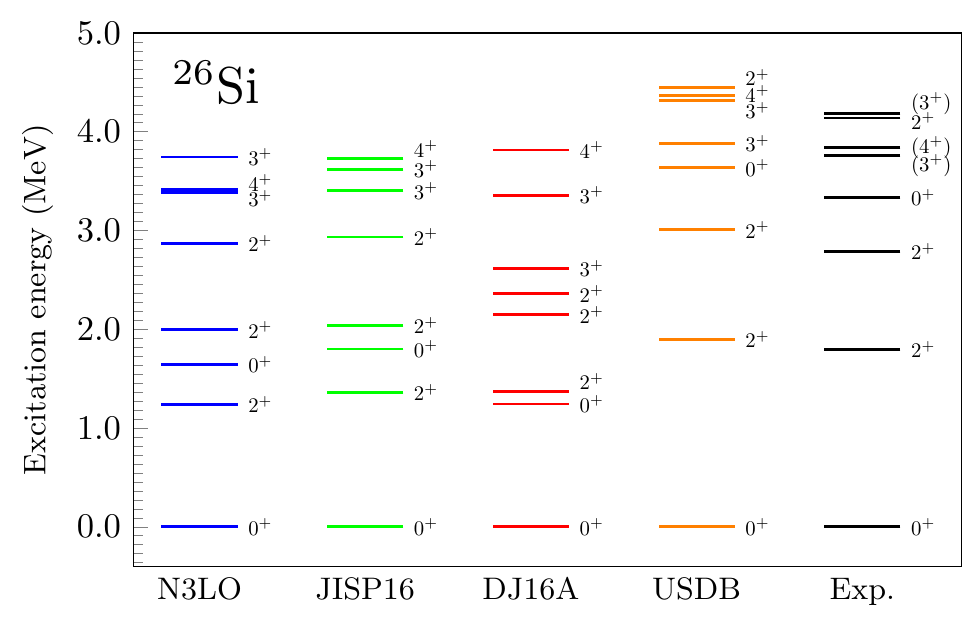}
	\includegraphics[width=7cm,height=5.5cm]{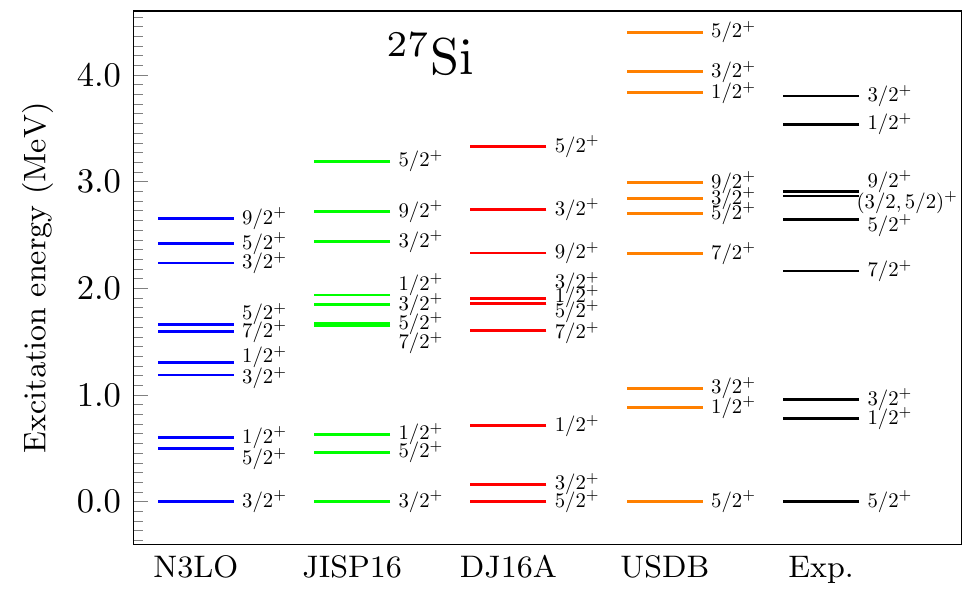}
	\includegraphics[width=7cm,height=5.5cm]{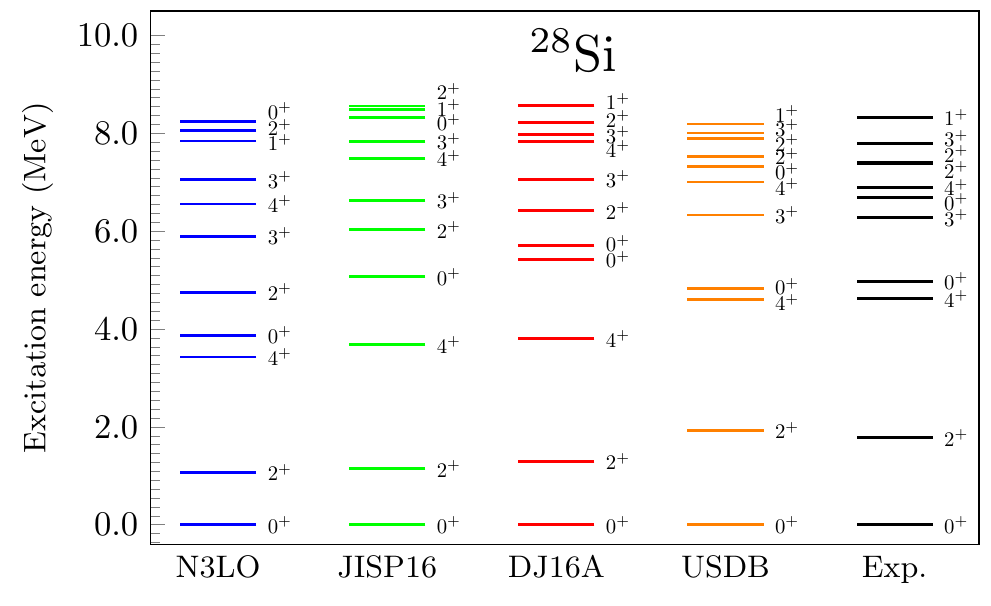}
	\caption{Low-lying energy spectra for Si isotopes in the range $A = 23 -28$. }
	\label{Si_isotopes_in_range_A23_32}
\end{figure*}

\begin{figure*}	
	\includegraphics[width=7cm,height=5.5cm]{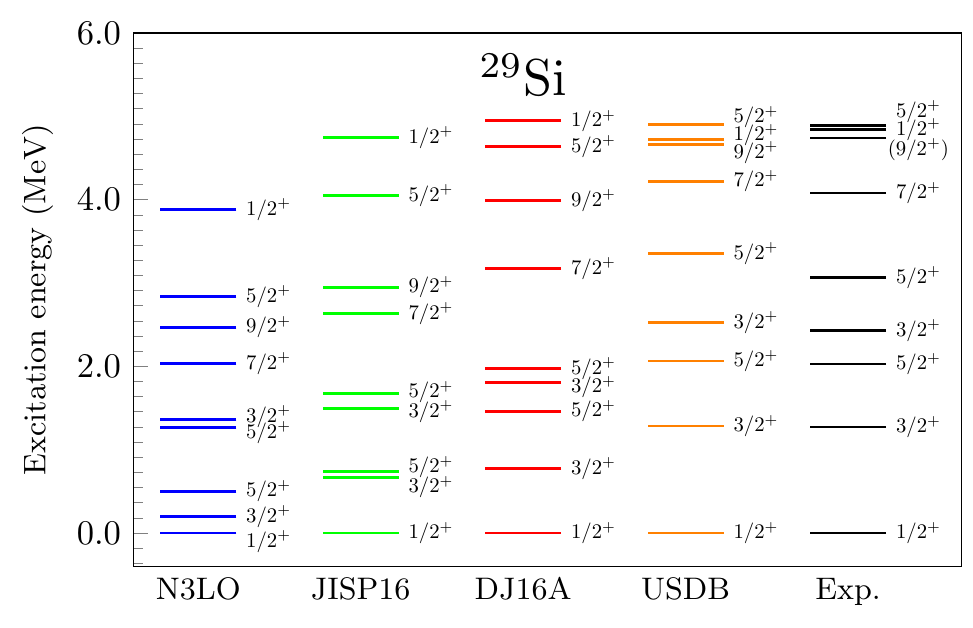}
	\includegraphics[width=7cm,height=5.5cm]{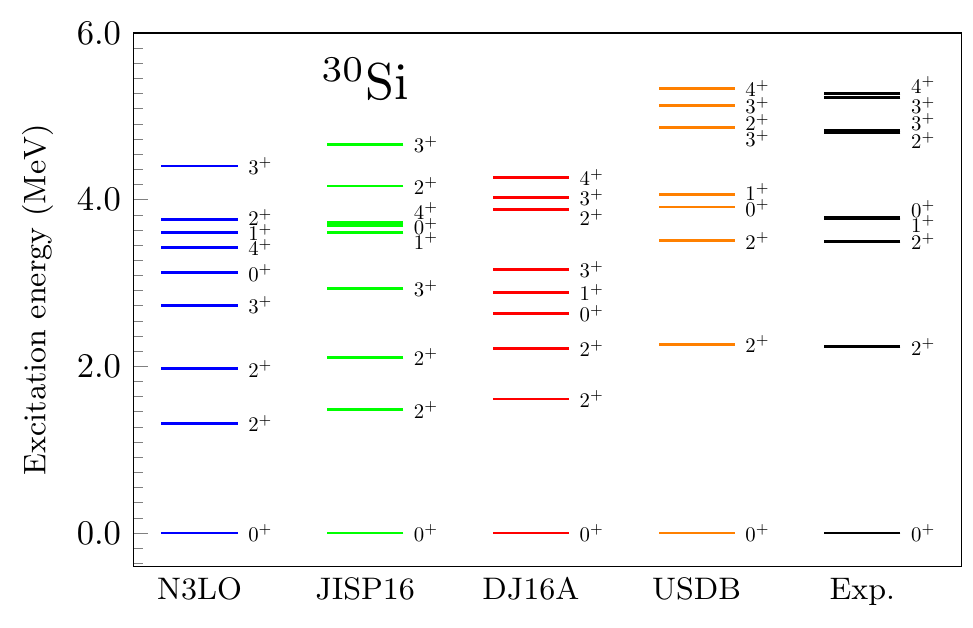}
	\includegraphics[width=7cm,height=5.5cm]{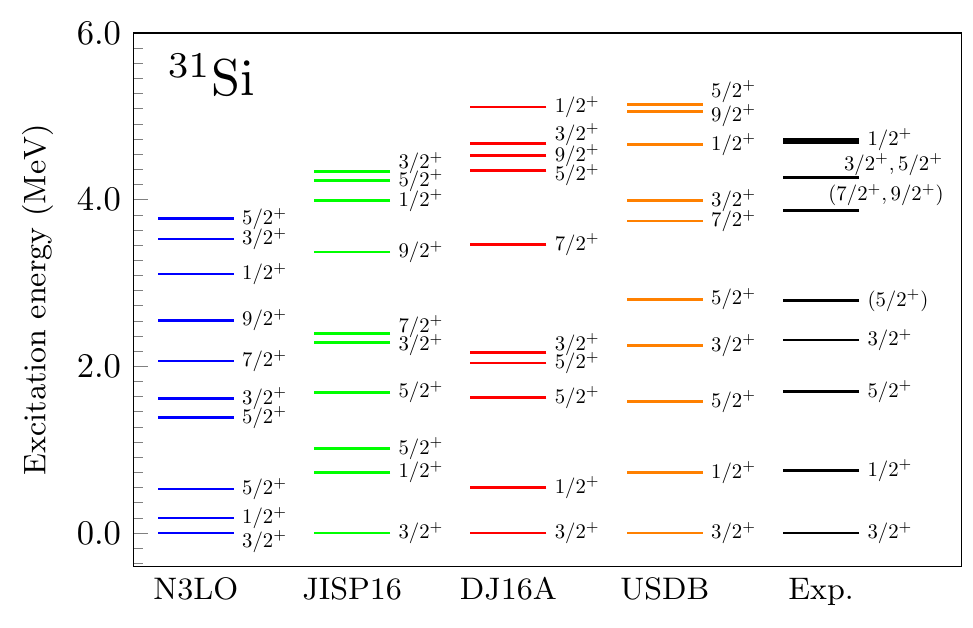}
	\includegraphics[width=7cm,height=5.5cm]{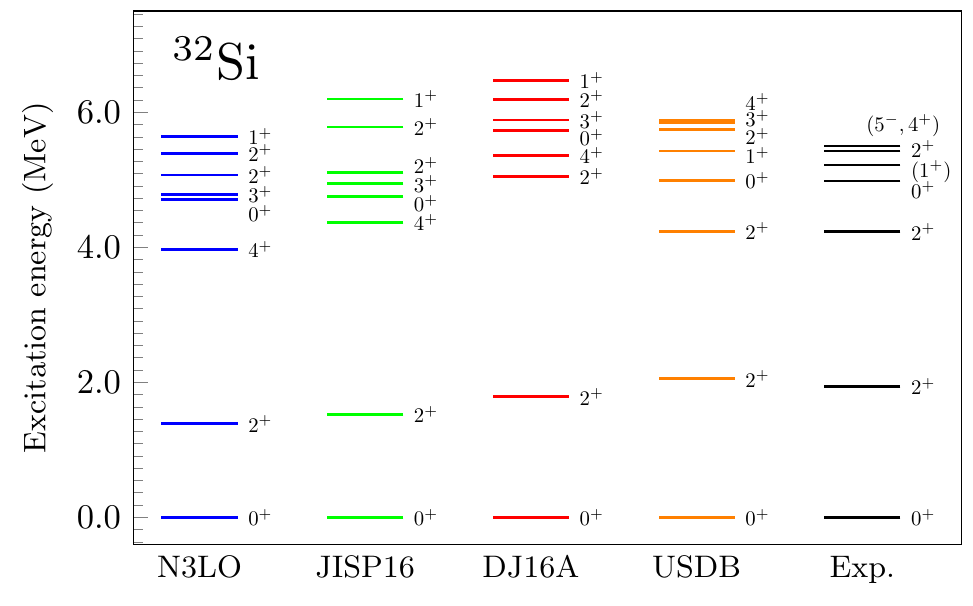}
	\includegraphics[width=7cm,height=5.5cm]{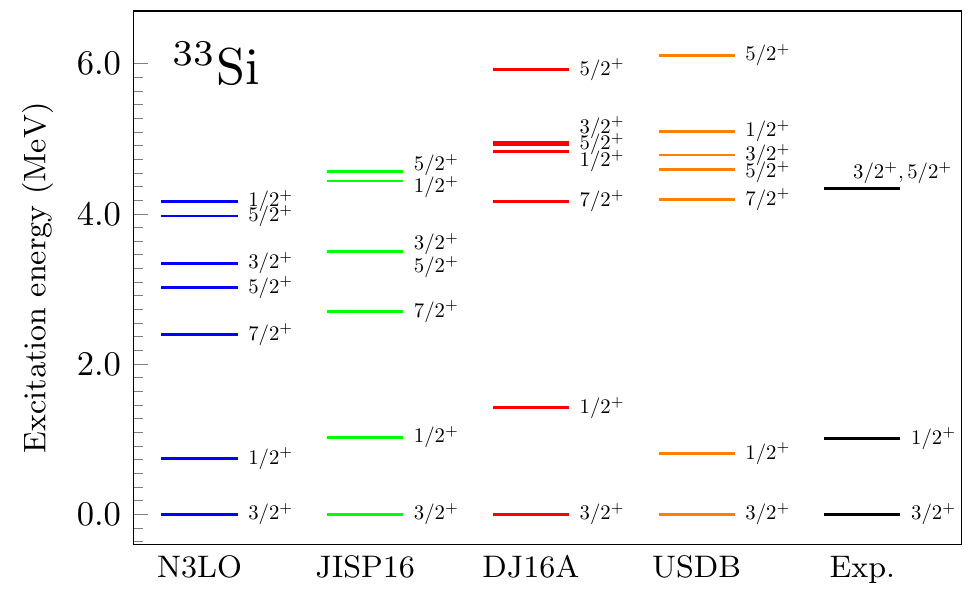}
	\includegraphics[width=7cm,height=5.5cm]{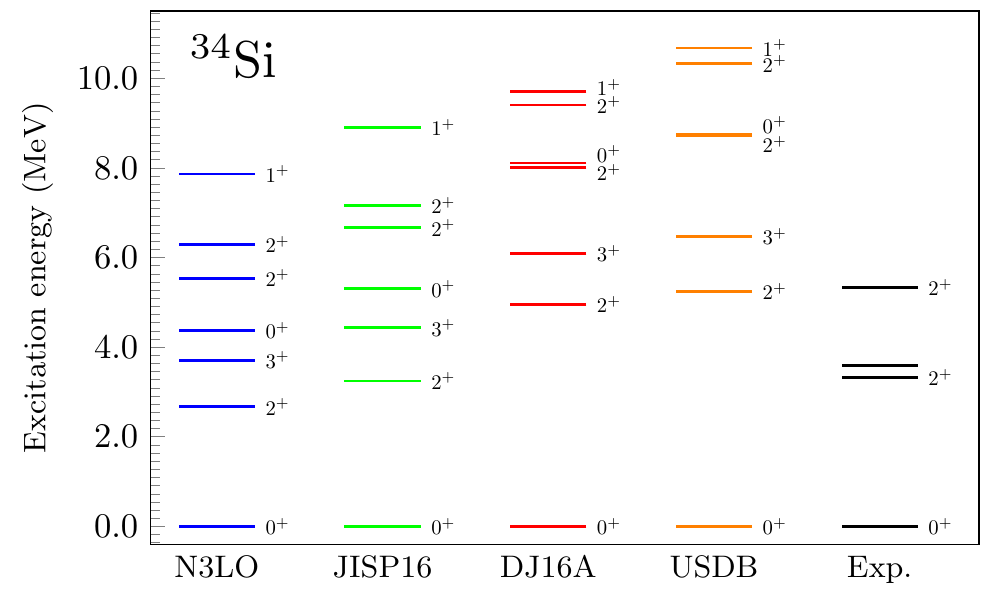}
	\caption{Low-lying energy spectra for Si isotopes in the range $A = 29 -34$. }
	\label{Si_isotopes_in_range_A33_34}
\end{figure*}

\begin{figure*}
	\includegraphics[width=7cm,height=5.5cm]{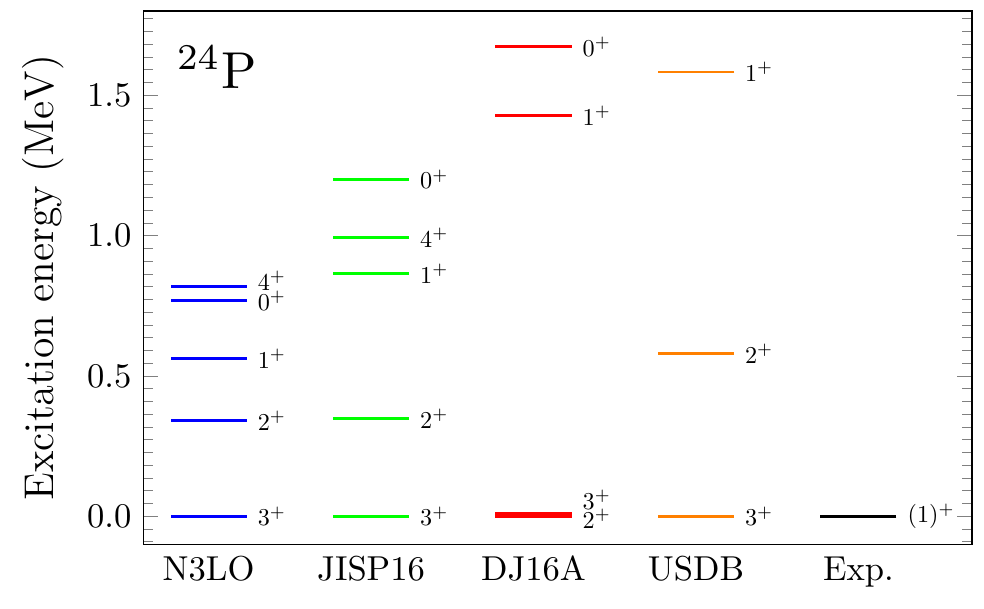}
	\includegraphics[width=7cm,height=5.5cm]{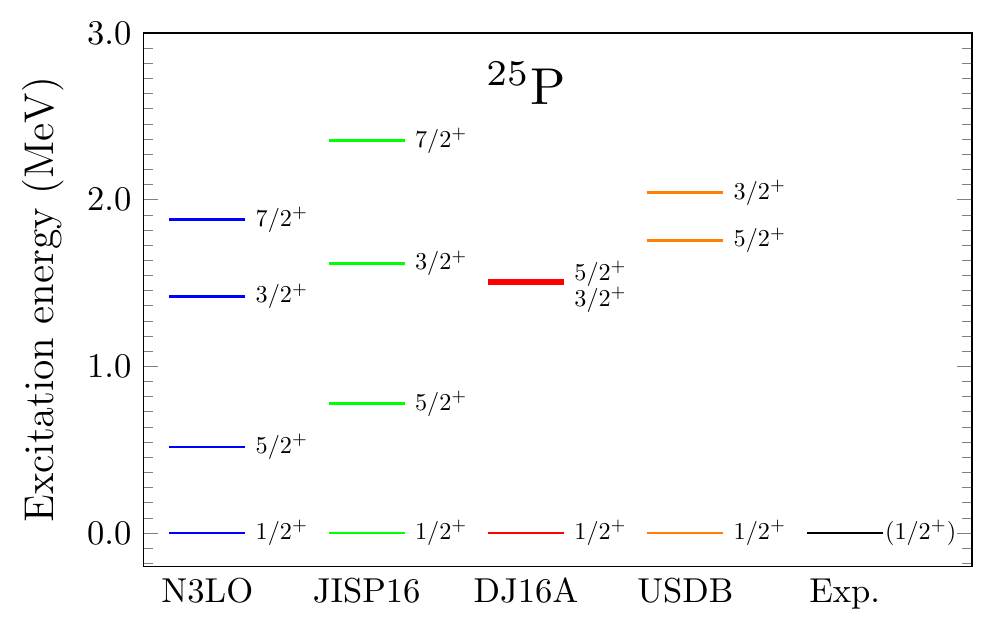}
	\includegraphics[width=7cm,height=5.5cm]{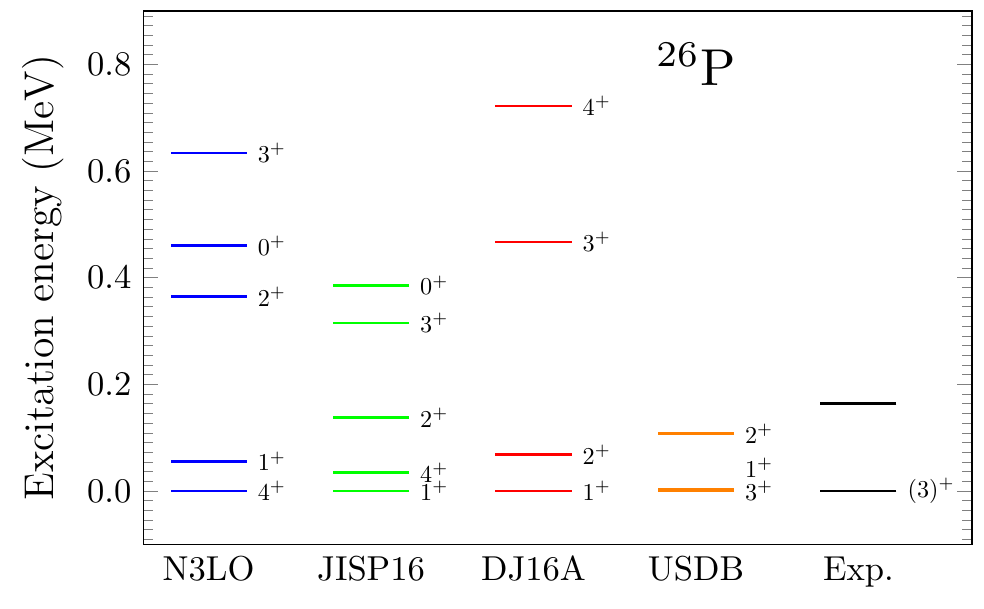}
	\includegraphics[width=7cm,height=5.5cm]{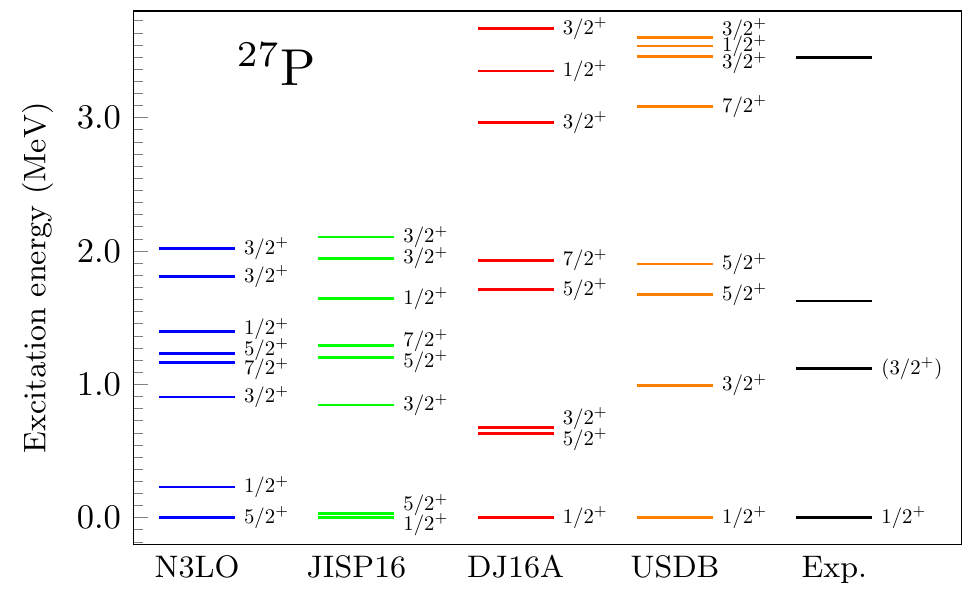}
	\includegraphics[width=7cm,height=5.5cm]{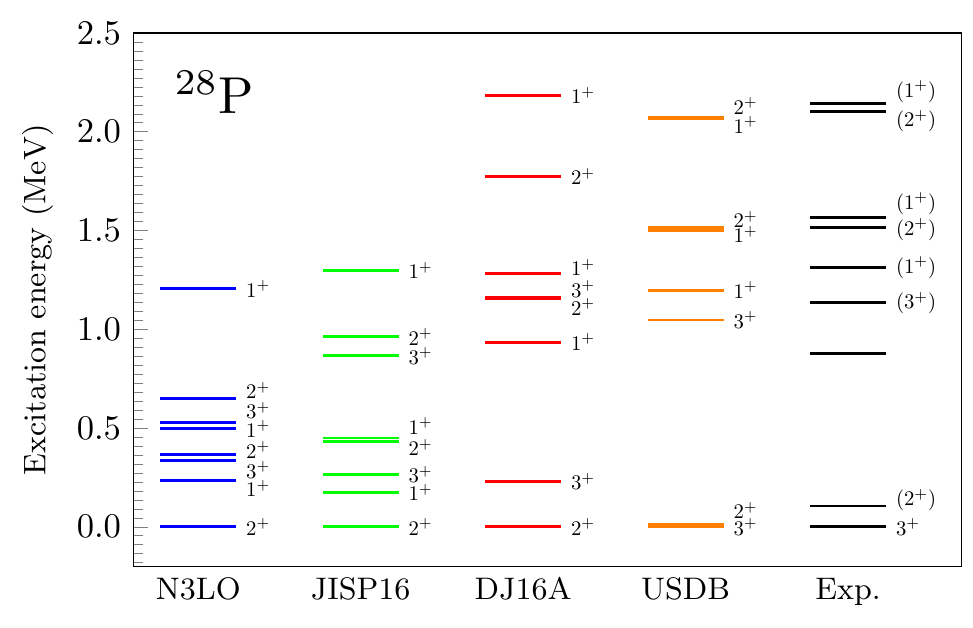}
	\includegraphics[width=7cm,height=5.5cm]{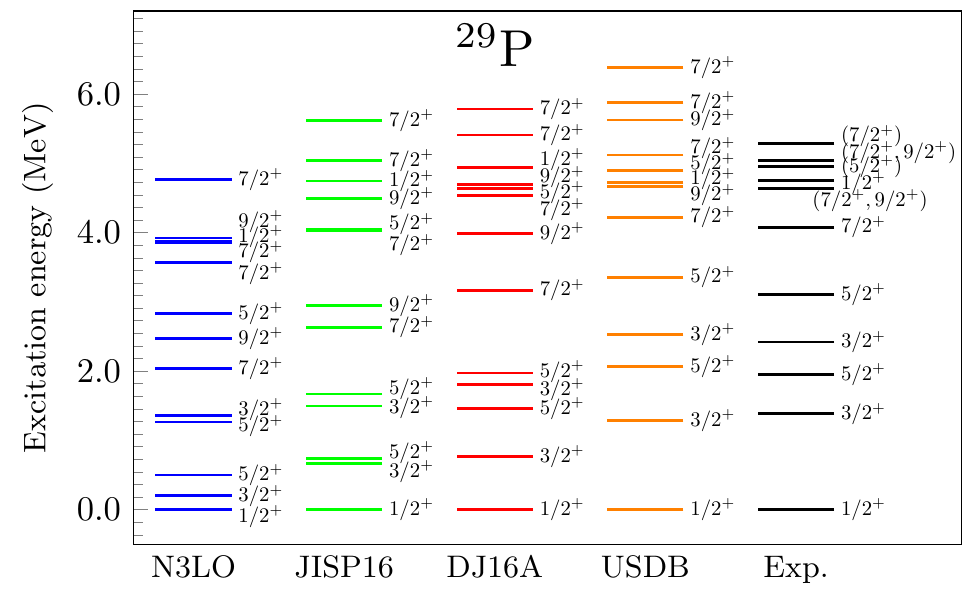}
	\caption{Low-lying energy spectra for P isotopes in the range $A = 24 -29$.}
	\label{P_isotopes_in_range_A24_33}
\end{figure*}

\begin{figure*}
	\includegraphics[width=7cm,height=5.5cm]{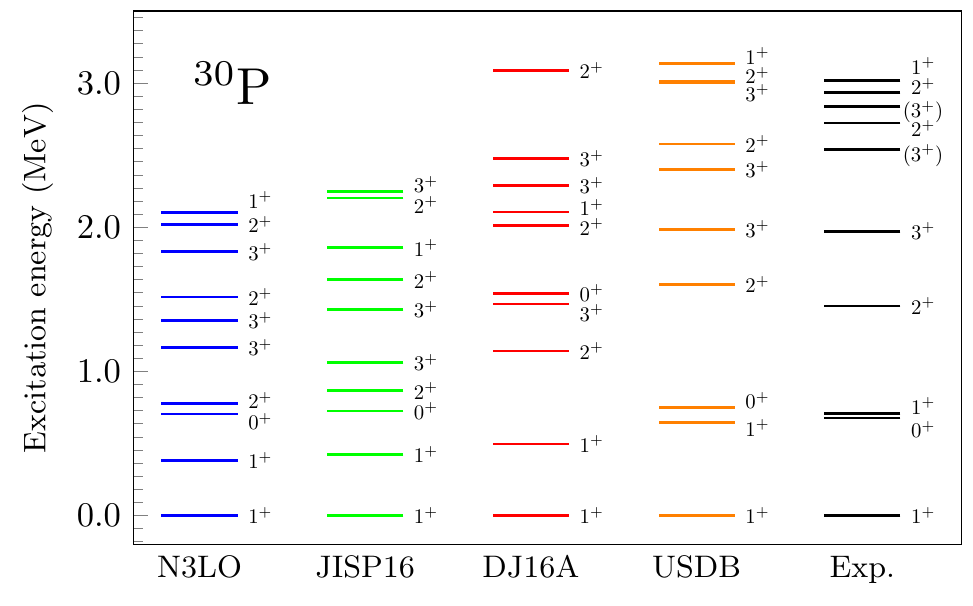}
	\includegraphics[width=7cm,height=5.5cm]{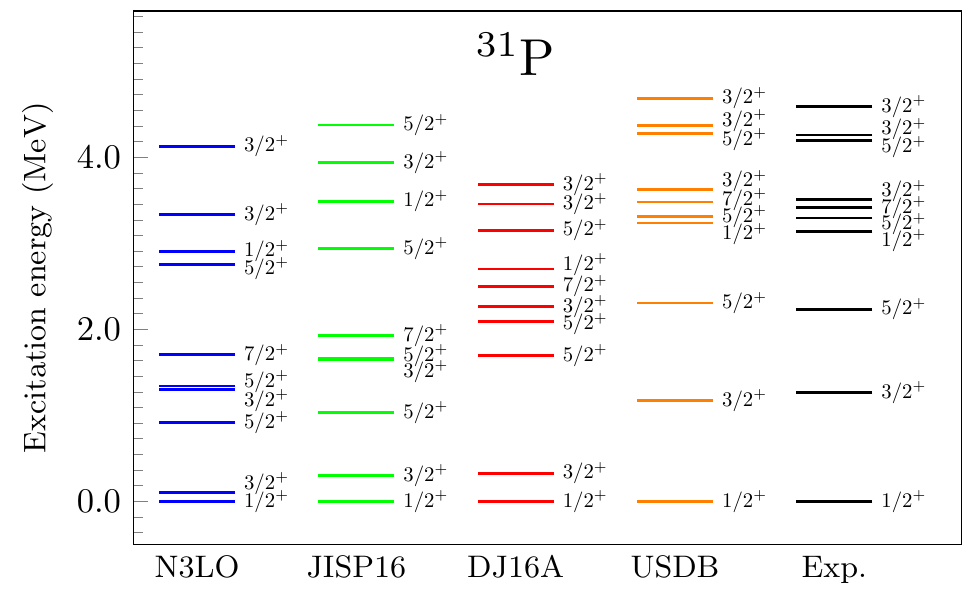}
	\includegraphics[width=7cm,height=5.5cm]{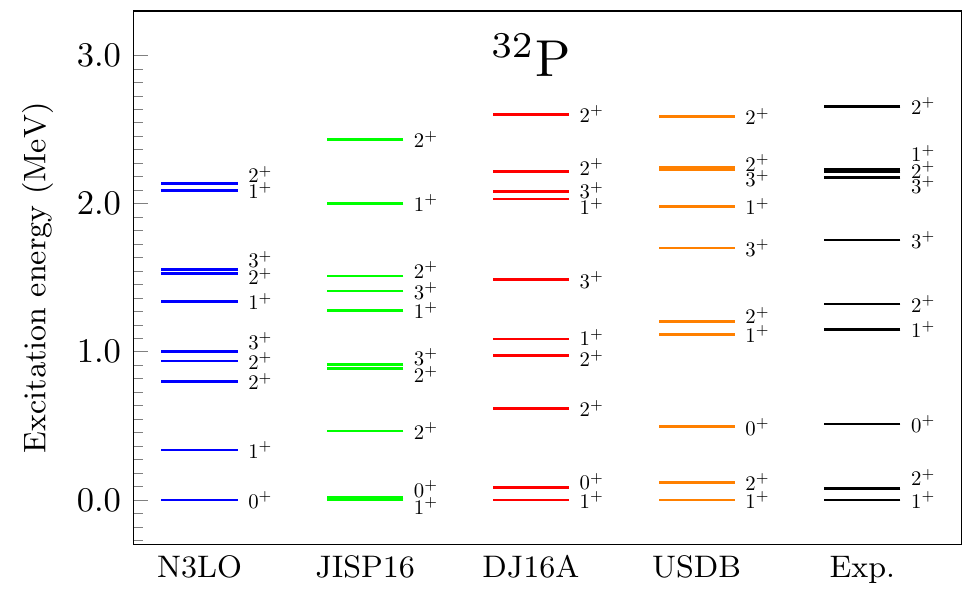}
	\includegraphics[width=7cm,height=5.5cm]{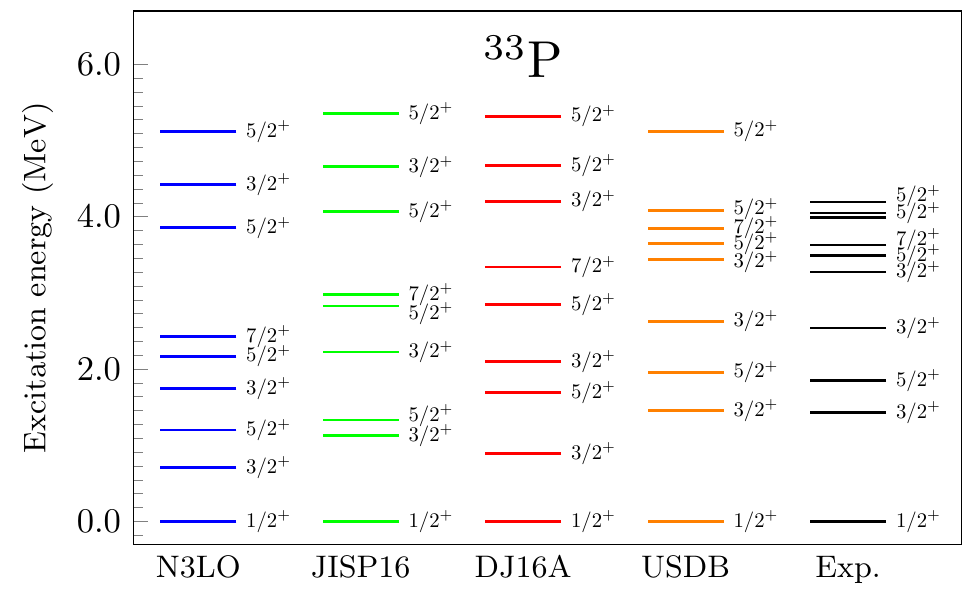}
	\includegraphics[width=7cm,height=5.5cm]{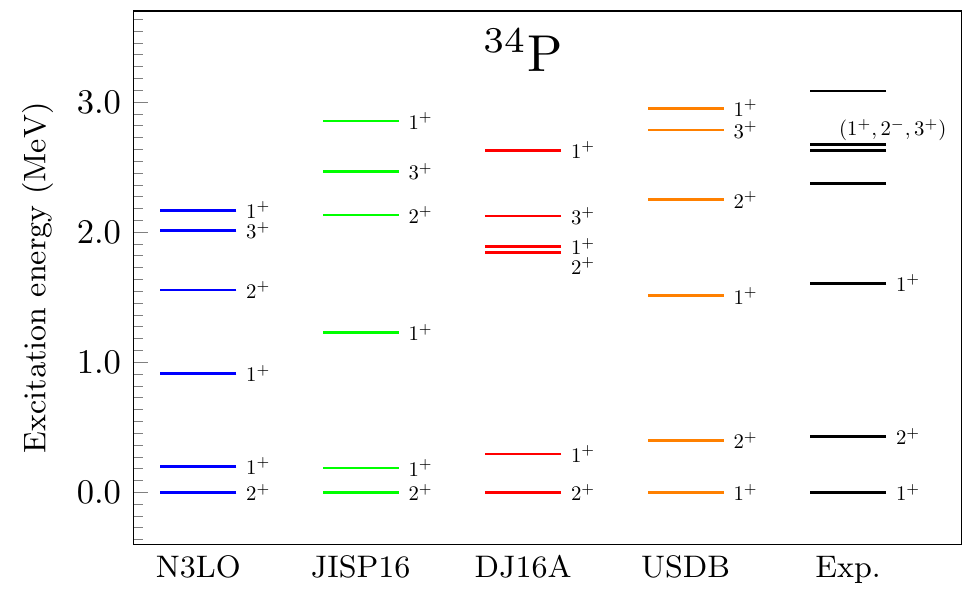}
	\includegraphics[width=7cm,height=5.5cm]{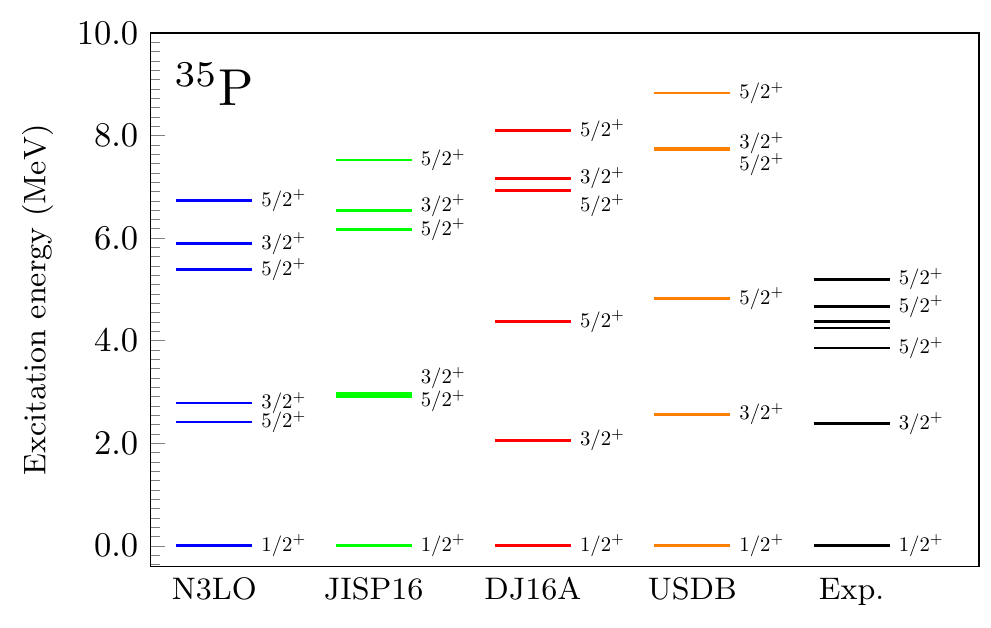}
	\caption{Low-lying energy spectra for P isotopes in the range $A = 30 -35$.}
	\label{P_isotopes_in_range_A34_35}
\end{figure*}

We have presented a comprehensive nuclear shell model results for $^{23-34}$Si and $^{24-35}$P isotopes using N3LO, JISP16 and DJ16A interactions. The empirical USDB calculations have also been performed for comparison. Here, we have shown the low-lying energy spectra of Si isotopes with $A = 23-34$ in Figs. \ref{Si_isotopes_in_range_A23_32}-\ref{Si_isotopes_in_range_A33_34} and P isotopes with $A = 24-35$ in Figs. \ref{P_isotopes_in_range_A24_33}-\ref{P_isotopes_in_range_A34_35}. Experimental spectrum \cite{NNDC} is shown in the last column of each figure.

\subsection{\bf Si isotopes}

All microscopic interactions confirm the tentative experimental spin of g.s. as 5/2\textsuperscript{+} for \textsuperscript{23}Si with dominant configuration of $\ket{\pi (d_{5/2}^6) \otimes \nu (d_{5/2}^1)}$, except for N3LO. In \textsuperscript{24}Si, correct g.s. 0\textsuperscript{+} is obtained with each interaction. For first excited state 2\textsuperscript{+}, DJ16A interaction gives 1.786 MeV excitation energy, while USDB interaction predicts this state to be at 2.111 MeV. Experimental value is 1.879 MeV which indicates that DJ16A interaction provides better agreement for 2\textsuperscript{+} state of \textsuperscript{24}Si, with the experiment. 
In \textsuperscript{25}Si, only g.s. spin-parity is measured experimentally and spin-parity of  excited states are unknown. Microscopic interactions are unable to reproduce spin of the g.s.
In case of \textsuperscript{26}Si, we observe from the spectrum that microscopic interactions produce $0_2^+$ state at small excitation energies ($\sim$ 1.2 - 1.6 MeV) in comparison with the experimental value (3.336 MeV). The $s_{1/2}$ proton orbital occupancy increases with excitation energy according to the USDB calculations as 0.58 for $0_1^+$, 0.68 for $2_1^+$, 1.03 for $2_2^+$ and with DJ16A as 0.84 for $0_1^+$, 0.92 for $2_1^+$, 1.05 for $2_2^+$. For \textsuperscript{27}Si, correct g.s. spin is reproduced with microscopic DJ16A only, which has the configuration of $\ket{\pi (d_{5/2}^6) \otimes \nu (d_{5/2}^5)}$ with probability of 16.42\%. USDB interaction also predicts $5/2^+$ as the g.s. of \textsuperscript{27}Si with the same configuration as found in case of DJ16A interaction with 27.36\% probability. 
The IM-SRG and CCEI give the major configurations in wave functions from $\ket{\pi (d_{5/2}^6) \otimes \nu (d_{5/2}^5)}$ and $\ket{\pi (d_{5/2}^5 s_{1/2}^1 ) \otimes \nu (d_{5/2}^4 s_{1/2}^1 )}$ with probabilities of 9.27\% and 13.24\%, respectively, as reported in Ref. \cite{PhysRevC.96.024316}.
With DJ16A, energy of $1/2_1^+$ is 0.711 MeV which is close to the experimental value of 0.781 MeV. The energy spectra with N3LO interaction is compressed as compared to the other interactions. 

$^{28}$Si and $^{29}$Si have been studied in \cite{PhysRevC.100.054329}, where, small excitation energies of $2_1^+$ and $4_1^+$ due to small $N = 14$ shell gap, and configurations of $0_{1}^+$ and $0_{2}^+$ states for $^{28}$Si have been described. We have added some excited states with all four interactions and have also studied spectroscopic properties (in Section VI). \textsuperscript{28}Si nucleus is the closed subshell nucleus which contains 6 valence protons and 6 valence neutrons. We obtain correct ordering of the states up to 0\textsubscript{2}\textsuperscript{+} in \textsuperscript{28}Si with each interaction. Energy difference between g.s. and first excited state is 1.068, 1.145, 1.296 and 1.932 MeV with N3LO, JISP16, DJ16A and USDB interactions, respectively, and corresponding experimental energy difference is 1.779 MeV. Similarly, excitation energy  for $4^+_1$ state approaches the experimental value as we go from N3LO to JISP16 to DJ16A. Excitation energies of all states increase from N3LO to DJ16A for \textsuperscript{29}Si and, hence, we see remarkable good  agreement between DJ16A and experimental energies. Also, a state at 4.741 MeV is confirmed to be $9/2^+$ with each interaction.
The g.s. of \textsuperscript{30}Si corresponding to the DJ16A interaction is dominated by the $N = 16$ closed subshell configuration $\ket{\pi (d_{5/2}^6) \otimes \nu (d_{5/2}^6 s_{1/2}^2)}$ with 24.39\% probability and the USDB interaction also  generates the same configuration with 27.54\% of probability. 
The $0_2^+$ has the similar configuration as g.s. for DJ16A interaction with 16.65\% of probability, while this state is populated using USDB interaction with $\ket{\pi (d_{5/2}^6) \otimes \nu (d_{5/2}^6 d_{3/2}^2)}$ and 
$\ket{\pi (d_{5/2}^6) \otimes \nu (d_{5/2}^6 s_{1/2}^2)}$ configurations with probabilities of 30.81\% and 17.44\%, respectively. 
We observe quite small improvement in the energies of 2\textsubscript{1}\textsuperscript{+} and  2\textsubscript{2}\textsuperscript{+} with DJ16A. We obtain correct ordering up to $5/2^+_1$ with each of the interactions in case of \textsuperscript{31}Si.

For \textsuperscript{32}Si, energy of first excited state is reasonably improved with DJ16A interaction (Fig.\ref{Si_isotopes_in_range_A33_34}). The $2^+_1$ and $2^+_2$ states have the configuration of \\ $\ket{\pi (d_{5/2}^6) \otimes \nu (d_{5/2}^6 s_{1/2}^2 d_{3/2}^2)}$ with probabilities of 17.98 and 48.62\%, respectively, using DJ16A interaction. The same configuration is obtained by USDB interaction with 20.71 and 48.78\% probabilities. We have calculated energy states with spin up to $7/2^+$ for \textsuperscript{33}Si. We can see from the spectrum  of \textsuperscript{33}Si in Fig. \ref{Si_isotopes_in_range_A33_34} that there is a large energy gap between $1/2^+$ and $7/2^+$ states for DJ16A and USDB interactions while this difference is small in case of N3LO and JISP16 interactions. Experimentally, a state is measured at 4.341 MeV with spins $3/2^+$ and $5/2^+$. So, we have computed both spin states with all four interactions. In \textsuperscript{34}Si, energy spectra obtained with DJ16A and USDB interactions are very similar. The \textsuperscript{34}Si has completely filled neutron $sd$ orbitals. For g.s., proton configuration with all four interactions is $\ket{\pi (d_{5/2}^6)}$. For the $2^+_1$ and $2^+_2$, the  configurations are $\ket{\pi (d_{5/2}^5 s_{1/2}^1)}$ and $\ket{\pi (d_{5/2}^5 d_{3/2}^1)}$ with probabilities of 89.71 and 88.77\%, respectively, corresponding to DJ16A interaction. The same configuration is achieved by USDB interaction for above mentioned states with 87.61 and 89.84\% probabilities, respectively.

When we look at the excitation energy of $2^+_1$ state for even-even Si isotopes with neutrons $N = 10 - 20$, we found that the USDB interaction gives better agreement with the experimental data for Si isotopes, except for $^{34}$Si. We can clearly observe from the experimental data that there is a shell gap at $N = 20$ for Si isotopes. For \textsuperscript{34}Si, DJ16A and USDB interactions produce large excitation energy for $2^+$ state compared to that obtained from the experiment by excitation of a proton from $d_{5/2}$ to $s_{1/2}$ orbit which shows a large subshell gap at $Z=14$, explained in the next section. Hence, $^{34}$Si is a doubly magic nucleus with a major closed neutron shell $N=20$ and a closed proton subshell $Z=14$. To reproduce the excitation energy of $2^+$, inclusion of neutron excitations from the $sd$-shell to $pf$-shell across the $N=20$ shell gap is needed as suggested in Ref.~\cite{Kaneko}.

\subsection{\bf P isotopes}

$^{25}$P is the least stable isotope while $^{31}$P is the most stable. For $^{24}$P and $^{25}$P, only g.s. is experimentally measured with unconfirmed spin-parity $(1^+)$ and $(1/2^{+})$, respectively. Nuclear shell model calculations with all four interactions confirm g.s. to be $1/2^{+}$ for $^{25}$P. The $^{26}$P has g.s. with tentative spin $(3)^{+}$, experimentally. The USDB interaction provides confirmation to g.s. spin, while other interactions fail to do so. In case of $^{28}$P, spin-parity of g.s. is $3^{+}$ experimentally, which is well reproduced by the empirical interaction, whereas microscopic interactions predict the g.s. to be $2^{+}$. First excited state has the spin of $(2^{+})$, experimentally, which is 105.6 keV above the g.s. The energy difference between $2^{+}$ and $3^{+}$ states is  337, 266 and 229 keV corresponding to N3LO, JISP16 and DJ16A interactions, respectively. With USDB interaction, $2^{+}$ is 13 keV above the $3^{+}$ state. Analysing the configurations of $3^{+}$ and $2^{+}$ states, it seems that both states have the same configuration of $\ket{\pi (d_{5/2}^6 s_{1/2}^1) \otimes \nu (d_{5/2}^5)}$ with USDB and DJ16A interactions.  It is observed that the energy spectra of N3LO and JISP16 interactions are quite compressed. The first $1^{+}$ state with the configuration $\ket{\pi (d_{5/2}^6 s_{1/2}^1) \otimes \nu (d_{5/2}^4 s_{1/2}^1)}$ obtained from the USDB and DJ16A interactions is found at higher energy in the spectrum than N3LO and JISP16. 

In $^{29}$P, dominant configuration of excited state $3/2^{+}$ comes from the excitation of proton from $s_{1/2}$ to $d_{3/2}$ with probabilities 32.50\% and 15.14\% using USDB and DJ16A interactions, respectively. Other interactions produce this configuration with very small probabilities.  Energy of excited states $3/2^{+}$ and $5/2^{+}$ is best reproduced with USDB followed by DJ16A for $^{31}$P. 
Experimentally, the g.s. spin for odd-P isotopes is $1/2^{+}$ and the first excited state is $3/2^{+}$ except in  case of $^{25}$P, for which, first excited state is not measured yet. DJ16A and USDB interactions reproduce the g.s. and first excited state correctly for odd-P isotopes, except for $^{27}$P. For $^{27}$P, DJ16A yields the first excited state as $5/2^{+}$. The proton configurations with USDB and DJ16A interactions are  $\ket{\pi (d_{5/2}^6 s_{1/2}^1)}$ for g.s. and $\ket{\pi (d_{5/2}^6 d_{3/2}^1)}$ for the first excited state with maximum probabilities. As neutrons increase from $A=25$ onwards, they sequentially occupy $d_{5/2}$, $s_{1/2}$ and $d_{3/2}$ orbitals for ground and first excited states with maximum probabilities, except for $^{31}$P. For $^{31}$P, the configurations of $3/2^{+}$ state are $\ket{\pi (d_{5/2}^6 d_{3/2}^1) \otimes\nu (d_{5/2}^6 d_{3/2}^2)}$ and $\ket {\pi (d_{5/2}^6 d_{3/2}^1) \otimes \nu (d_{5/2}^6 s_{1/2}^2)}$ for DJ16A interaction with probabilities of 10.15 and 9.29\%. With USDB interaction, the same configuration is obtained with 15.80 and 15.34\% probabilities.

$^{35}$P has one proton hole beyond $Z=16$ with $N=20$,  hence, lowest $3/2^+$ excited state is produced by the excitation of proton from $s_{1/2}$ to $d_{3/2}$ orbital. Large energy splitting between $3/2^+_1$ and $1/2^+_1$ at $N=20$ indicates a shell gap at $Z=16$, although this shell gap is smaller than shell gap observed at $Z=14$.

\begin{figure*}[h]
	\includegraphics[width=6cm]{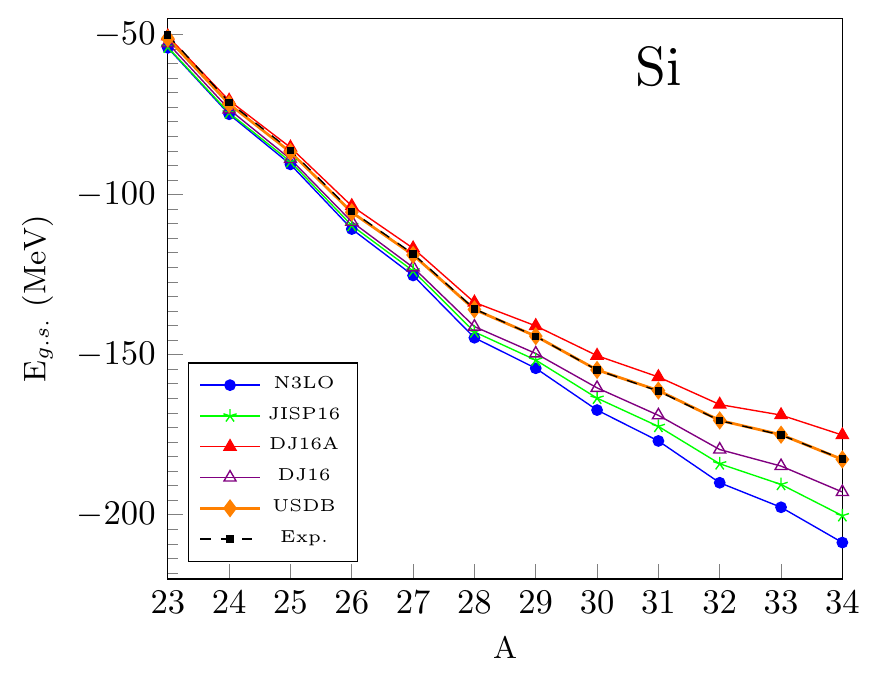}
	\includegraphics[width=6cm]{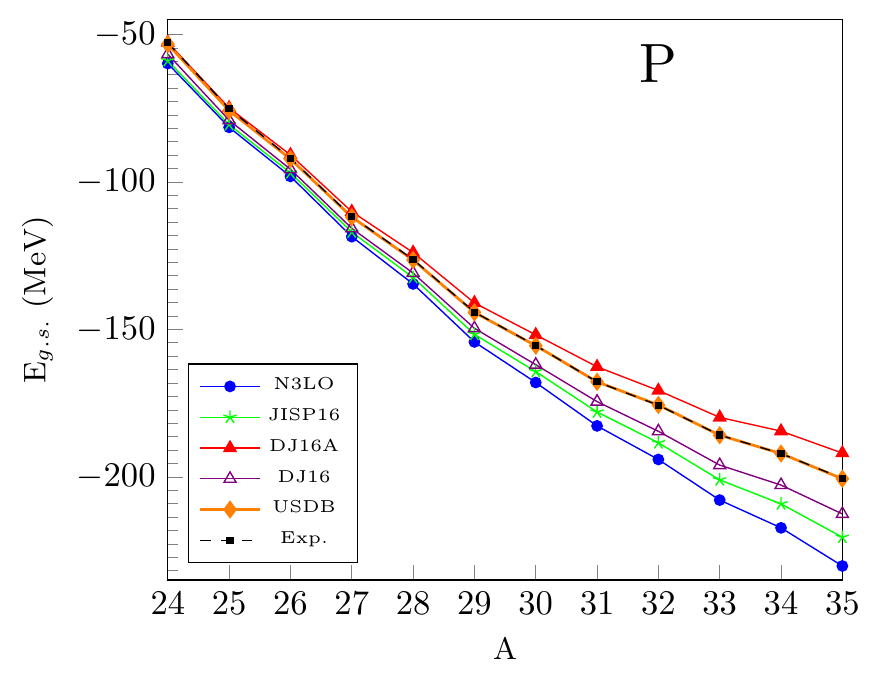}
	\caption{
		Comparison of experimental g.s. energy \cite{NNDC} with the shell model results for Si and P isotopes with respect to g.s energy of $^{16}$O.
	}
	\label{G.s. spectra for Si isotopes}
\end{figure*}

Energies of g.s. for Si and P isotopes relative to g.s. of $^{16}$O with mass number $A = 23-34$ and $A = 24-35$ are  depicted in Fig. \ref{G.s. spectra for Si isotopes}. As the number of neutrons in Si and P isotopes increases from $A = 23/24$ to $A = 34/35$, the energy difference between theory and experiment increases for all three microscopic interactions. The N3LO and JISP16 interactions overbind the g.s., while DJ16A interaction slightly underbinds the g.s. (which is discussed in Section V). 
The failure of the N3LO interaction is evident in Fig. 3 due to the missing 3N interaction. If the 3N interaction is included, the agreement would be improved. The  DJ16A and JISP16 interactions are fitted only up to $A=16$ nuclei which give rise to deficiencies in results of Si and P isotopes. The JISP16 cannot be applied beyond $^{16}$O, which explains the failure of this interaction to reproduce the g.s. energies in Fig. 3. The failure of JISP16 is not due to this interaction, but is due to its application beyond the limit, hence, the discrepancy is anticipated. The DJ16 interaction overbinds the g.s. energy of Si and P isotopes. The DJ16A
results are closer to the experimental data than the DJ16 results.
DJ16A is monopole modified version of the DJ16 interaction, it produces better results for g.s. energies than the JISP16 due to the non-local addition as a replacement of 3N interaction. The DJ16A and JISP16 interactions yield quite similar results. They produce different results mainly for the g.s. energies. To get more accurate results with DJ16A, non-monopole modifications are needed for further tuning of TBMEs. It is also possible to resolve these discrepancies by the use of (i) primary effective Hamiltonian with larger NCSM $N_\mathrm{max}$ parameter, (ii) three-body components induced by the OLS transformation ($a$ = 3 cluster approximation), and (iii) three-nucleon interaction in the NCSM Hamiltonian in addition to realistic NN interaction. 

\section{Effective single-particle energies: Monopole properties}
\begin{figure*}
	\includegraphics[width=6.0cm]{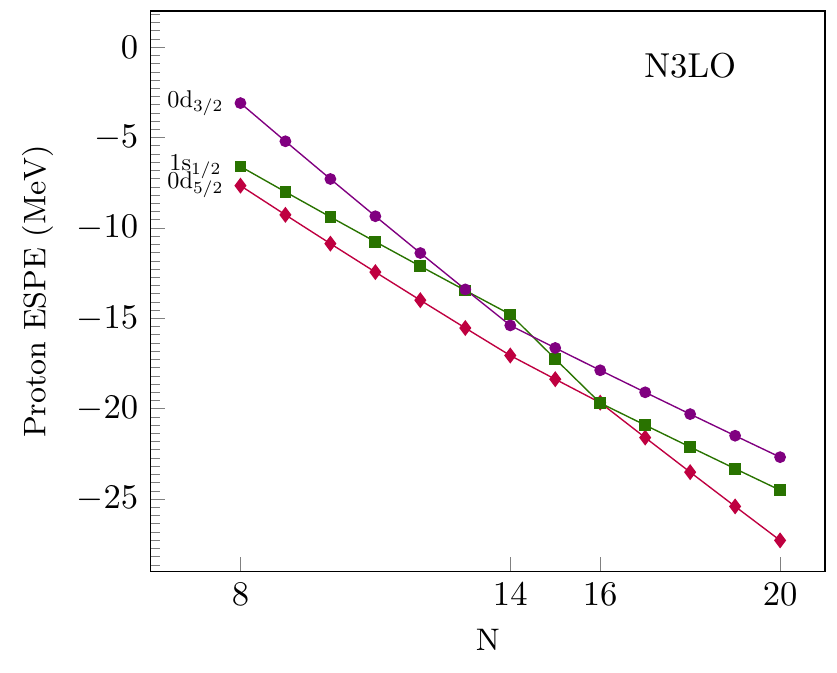}
	\includegraphics[width=6.0cm]{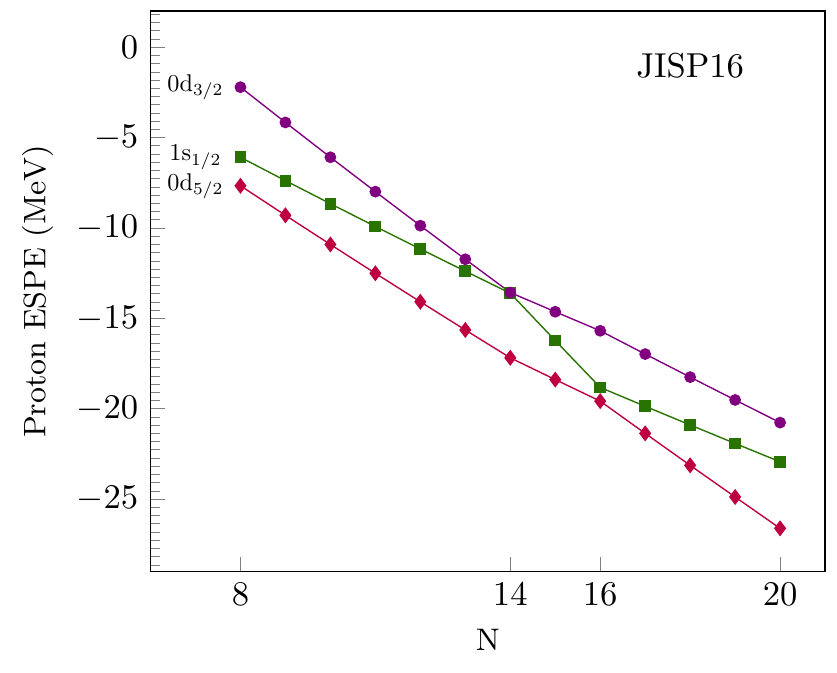}
	\includegraphics[width=6.0cm]{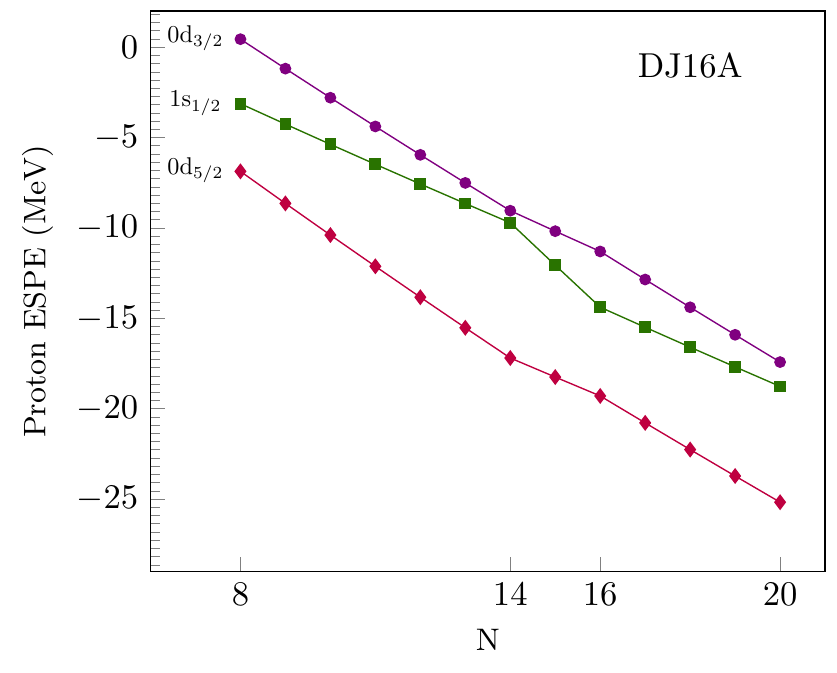}
	\includegraphics[width=6.0cm]{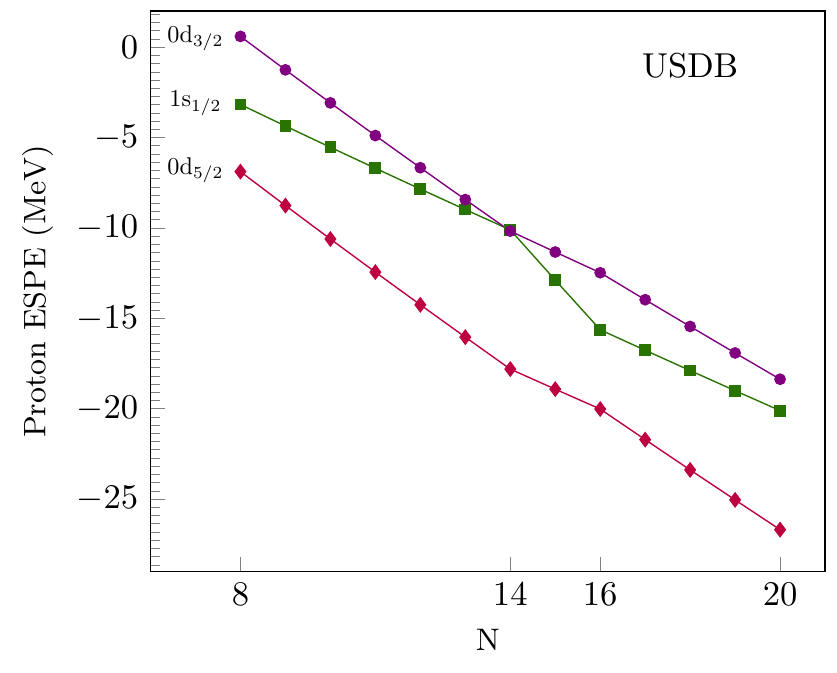}
	\caption{Proton effective single-particle energies of Si isotopes, calculated with the N3LO, JISP16, DJ16A and USDB interactions.}
	\label{monopole}
\end{figure*}
The Hamiltonian can be classified into monopole and multipole  (pairing, quadrupole-quadrupole correlations etc.) parts.
The monopole part \cite{BaFr64} of the Hamiltonian tells about the spherical nuclear mean fields that give information of filling of nucleons in the single particle states \cite{PoZu81}. It provides the position of the single particle orbitals which play important role in describing the evolution of shell gap. In this section, we have shown the proton effective single particle energies (ESPEs) \cite{Ot01,UmMu04,monopole1} corresponding to N3LO, JISP16, DJ16A and USDB interactions for Si isotopes in Fig. \ref{monopole}.
ESPEs describe how single particles energies of $sd$ orbitals vary as we increase  neutrons in the silicon nuclei.
In calculation of ESPEs,  monopole term of two-body interaction is needed which is given by
\begin{equation}
	V(j j^{'}) = \frac{\sum_{J}(2J+1)\bra{j j^{'};J}V\ket{j j^{'};J}}{\sum_{J}(2J+1)}.
\end{equation}
Monopole interaction shows an average effect between two nucleons in orbits $j$ and $j^{'}$.  This monopole interaction changes the single-particle energy and consequently the shell gaps as we add more nucleons. Thus, spin-tensor decomposition of the monopole interaction is carried out in next section to understand the origin of this shell evolution. Effective single-particle energy and role played by different components of the two-body interaction in the shell evolution had been discussed in several papers \cite{Tsunoda:2020gpt,monopole1,monopole,monopole2,monopole3,monopole4,monopole5,monopole6}.

Si is a mid shell nuclei, for which, configurations of states have contributions from both protons and neutrons.
From Fig. \ref{monopole}, we can see that ESPEs of orbitals $d_{5/2}$ and $s_{1/2}$ are very close to each other for N3LO interaction due to small difference between $V_{d_{5/2}d_{5/2}}^{pn}$ ($\bra{d_{5/2}d_{5/2}}V\ket{d_{5/2}d_{5/2}}$ ) and $V_{s_{1/2}d_{5/2}}^{pn}$ centroids at $^{28}$Si. ESPEs of orbitals $s_{1/2}$ and $d_{3/2}$ are inverted for N3LO, while for JISP16 and USDB interactions, these orbitals almost overlap in $^{28}$Si. 
We note that $V_{d_{5/2}d_{5/2}}^{pn}$ is less attractive, while $V_{d_{3/2}d_{5/2}}^{pn}$ is more attractive for N3LO and JISP16 than centroids for DJ16A and USDB, thus, proton $d_{5/2}-d_{3/2}$ spin-orbital splitting is smaller for N3LO and JISP16 than obtained from DJ16A and USDB at $N=14$. As mentioned in the previous section, there is high-lying $2_1^+$ state for $^{34}$Si predicted by DJ16A and USDB interactions, this high excitation energy is due to sufficiently large energy difference between $d_{5/2}$ and $s_{1/2}$ orbitals. So, it can be argued that there is a subshell closure at $Z=14$ for Si isotopes. This shell gap increases from N3LO to JISP16 to DJ16A to USDB interaction, thus, excitation energy for this state increases accordingly. It is also observed that the shell gap at $N=16$ between orbitals $s_{1/2}$ and $d_{3/2}$ is very small for microscopic N3LO interaction as compared to other interactions. 

\begin{figure*}[h]
\centering
	\includegraphics[width=8.5cm]{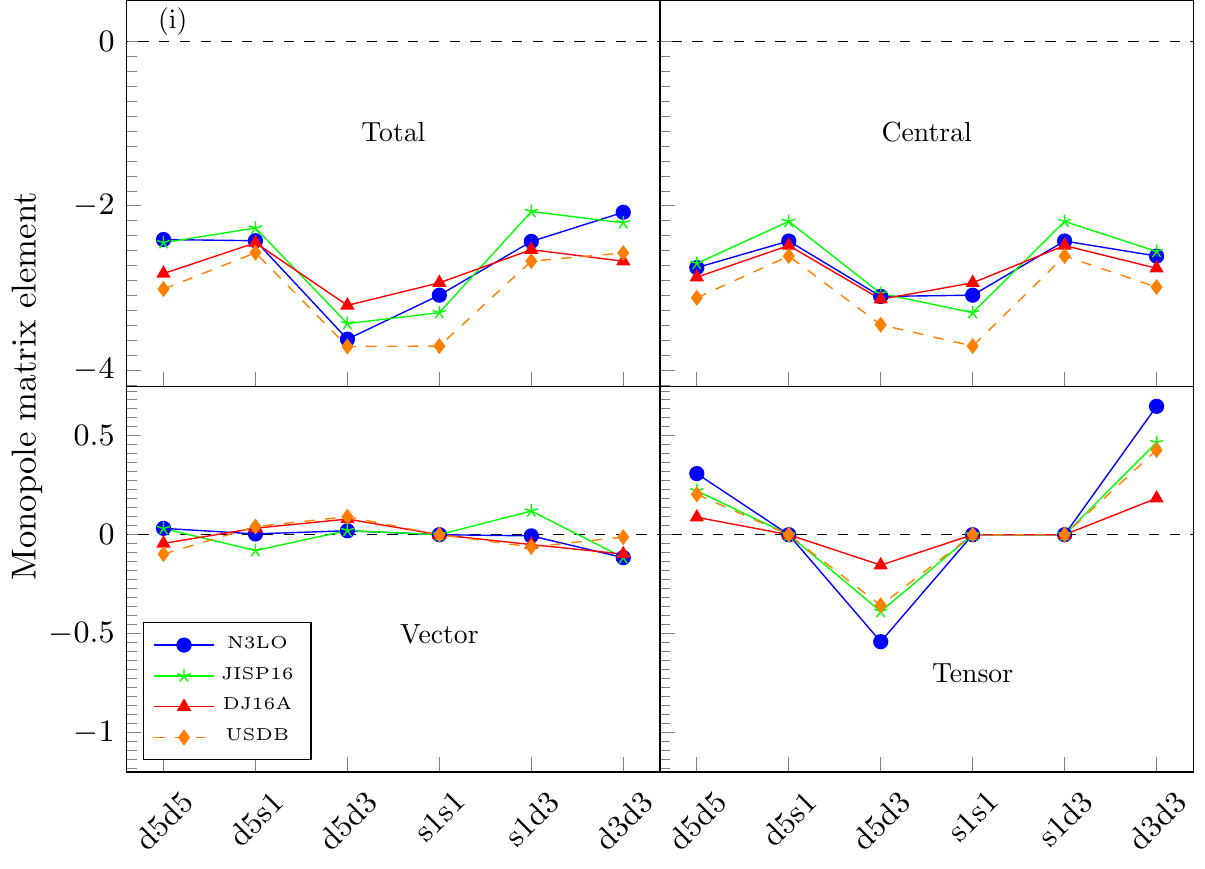}
	\includegraphics[width=8.5cm]{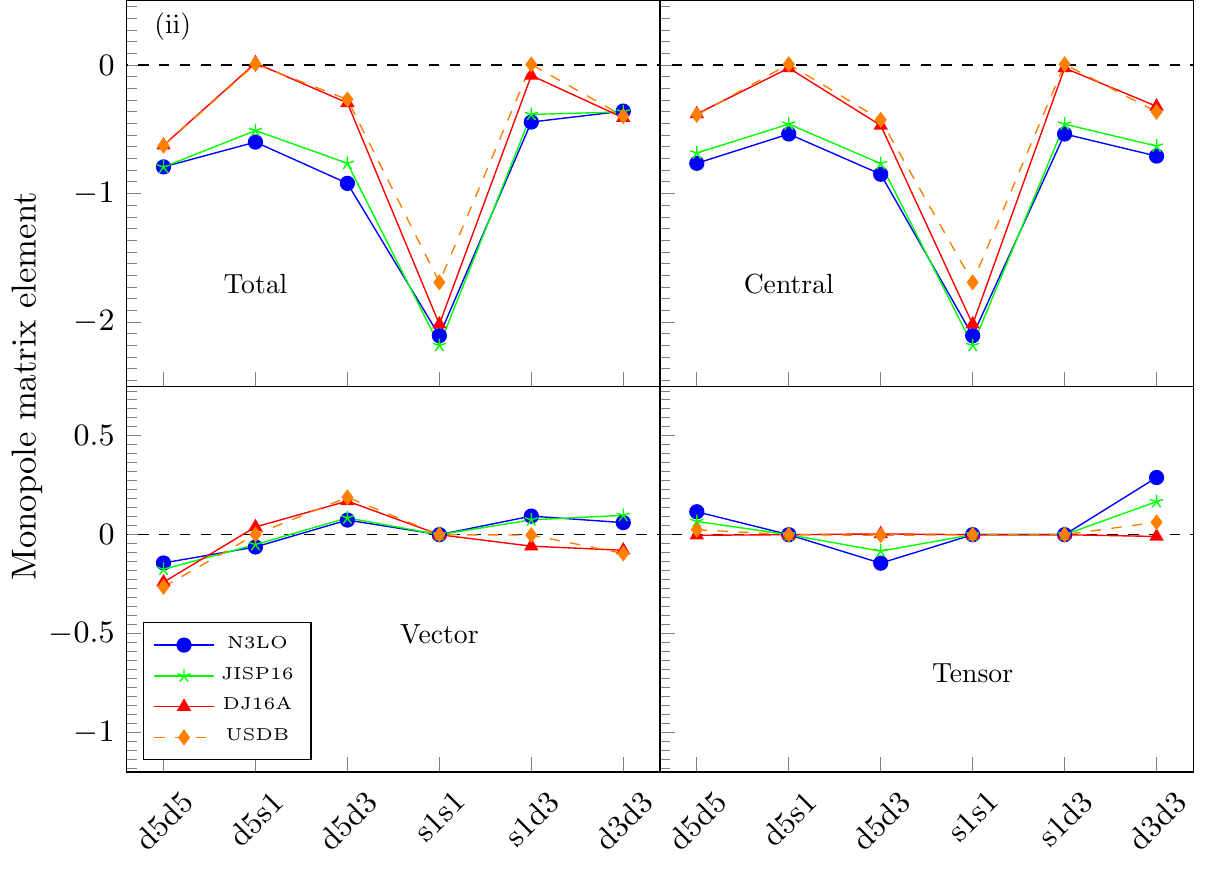}
	\caption{Monopole terms of total, central, vector and tensor forces for (i) T = 0 and (ii) T = 1 channels corresponding to N3LO, JISP16, DJ16A and USDB interactions.}
	\label{T0T1}
\end{figure*}
\section{Spin-Tensor decomposition of the two-body interaction}
To understand the contributions from different components of two-body interactions,
we have carried out spin-tensor decomposition \cite{Kir73,KlKn77}. A two-body interaction can be written as
\begin{equation}
	V = \sum_{k=0,1,2}(S^{(k)}.Q^{(k)}) = \sum_{k=0,1,2} V^{(k)},
\end{equation}

where, $k$ runs over 0, 1, 2. The term with $k=0$ indicates the central component, the term with $k=1$ the vector component and the term with $k=2$ the tensor component. $S^{(k)}$ are spherical tensors of rank $k$ in spin space, constructed for spin-1/2 nucleons and $Q^{(k)}$ are spherical tensors of rank $k$ in coordinate space.  The $1$ and ($\vec{\sigma_1}$.$\vec{\sigma_2}$) operators correspond to the spin part of the central component of the effective interaction. The spin part of the tensor component is given by $[\vec{\sigma_1}\times\vec{\sigma_2}]^{(2)}$.
The vector component includes symmetric spin-tensor forces, which has operator  $\vec{\sigma_1}$+$\vec{\sigma_2}$ and antisymmetric spin-tensor forces, which have operators $\vec{\sigma_1}$-$\vec{\sigma_2}$ and $[\vec{\sigma_1}\times\vec{\sigma_2}]^{(1)}$.

To determine the spin-tensor decomposition of the two-body interaction, firstly, one has to transform TBMEs to LS-coupling from jj-coupling as  follows

	\begin{equation}
		\begin{split}
			&\bra{n_a l_a n_b l_b: LS, JT}V\ket{n_c l_c n_d l_d: L^{'}S^{'}, JT} 
			= \sum_{j_a j_b} \frac{N_{l_a l_b}}{N_{j_a j_b}} \sqrt{(2L+1)(2S+1)(2 j_a +1)}  \\ 
			&	\times \sqrt{(2 j_b+1)}\gj{l_a}{l_b}{L}{1/2}{1/2}{S}{j_a}{j_b}{J}\sum_{j_c j_d} \frac{N_{l_c l_d}}{N_{j_c j_d}} \sqrt{(2L^{'}+1)(2S^{'}+1)(2 j_c +1)(2 j_d+1)} \\
			& \times\gj{l_c}{l_d}{L^{'}}{1/2}{1/2}{S^{'}}{j_c}{j_d}{J}
			\bra{n_a l_a j_a n_b l_b j_b: JT}V\ket{n_c l_c j_c n_d l_d j_d: JT},
		\end{split}
	\end{equation}

where, normalization factor $N_{l_a l_b}$ is $$N_{l_a l_b} = \frac{1}{\sqrt{2(1+\delta_{l_a l_b})}}.$$

In LS-coupling, matrix elements of each component of the two-body interaction are written as

	\begin{equation}
		\begin{split}
			&\bra{n_a l_a n_b l_b: LS, J^{'}T}V^{(k)}\ket{n_c l_c n_d l_d: L^{'}S^{'}, J^{'}T} 
			= (2k+1) (-1)^{J^{'}} \Gj{L}{S}{J^{'}}{S^{'}}{L^{'}}{k} \\
			&	 \times \sum_{J} (-1)^{J} (2J+1) \Gj{L}{S}{J}{S^{'}}{L^{'}}{k} \bra{n_a l_a n_b l_b: LS, JT}V\ket{n_c l_c n_d l_d: L^{'}S^{'}, JT}.
		\end{split}
	\end{equation}

In this section, we have studied the monopole matrix elements of the total, central, vector and tensor components for T = 0 and T = 1 channels for the N3LO, JISP16, DJ16A and USDB interactions, which are shown in Fig. \ref{T0T1}. Total monopole term of T = 0 channel is more attractive for all interactions as compared to T = 1 channel. We can see that the central forces play a dominant role in total two-body interaction. The  vector monopole terms of $V_{s_{1/2}s_{1/2}}$ are zero for T = 0 and T = 1 channels.
The tensor monopole terms of $V_{d_{5/2}s_{1/2}}$, $V_{s_{1/2}s_{1/2}}$ and $V_{s_{1/2}d_{3/2}}$ are zero corresponding to all interactions for both channels. 
For $V_{d_{5/2}d_{5/2}}^{T=0}$ centroid, central term is more attractive and tensor term is smaller for DJ16A and USDB than N3LO and JISP16. 
For the central component, the monopole matrix element of $V_{s_{1/2}s_{1/2}}$ for T = 0 channel of DJ16A interaction is less attractive in comparison with other interactions. Both T = 0 and T = 1 vector monopole terms are relatively flat for all interactions. 
The T = 0 tensor monopole term of $V_{d_{5/2}d_{5/2}}$, $V_{d_{5/2}d_{3/2}}$ and $V_{d_{3/2}d_{3/2}}$ are stronger than T = 1 tensor monopole.
The total, central, vector and tensor monopoles of $V_{d_{3/2}d_{3/2}}$ of $T=0$ channel for N3LO interaction are -2.079, -2.612, -0.116 and 0.649 MeV, which means tensor forces give significant contribution in total two-body interaction. Fig. \ref{T0T1} shows that the tensor force of DJ16A may be too weak because the N3LO is more fundamentally endorsed. Similar to Fig. \ref{monopole}, a figure was shown in Ref. \cite{PhysRevC.95.021304}, where the effective single-particle energies of \textit{ab initio} EEdf1 interaction appears to resemble another \textit{ab initio} interaction N3LO. The tensor force shifts the ESPE, and hence changes the separation between the orbitals.
The DJ16A T = 1 total monopoles are similar to those obtained with USDB interaction except for $V_{s_{1/2}s_{1/2}}$, for which, the former is attractive and latter is repulsive but with small variation.

It is noted that the DJ16A interaction underbinds the energies of g.s. for Si and P isotopes. From the above understanding of TBMEs of the interactions, we can say that this underbinding is related to the central part of T=0 channel which is more attractive for USDB interaction than DJ16A interaction and overbinding of N3LO and JISP16 results occur due to attractive $T=1$ centroids.

\section{ Electromagnetic properties}
In this section, we have presented reduced electric quadrupole transition strengths, electric quadrupole and magnetic dipole moments for Si and P isotopes. We have shown a comparison between theoretical and available experimental data.

\begin{table*}
	\caption{\label{B(E2)}Comparison of calculated reduced electric quadrupole transition strengths using different interactions of Si and P nuclei, with the experimental data \cite{NNDC}. The B(E2) values are in $e^2$fm$^4$.}
	\begin{tabular}{cccccccc}
		\vspace{-2.8mm}\\
		\hline
		\hline 
		Nuclei & A	&	$J_i^{\pi} \rightarrow J_f^{\pi}$	&	N3LO	&	JISP16	&	DJ16A		&	USDB&	Exp.	\T\B \\
		\hline
		Si	&26	&	$2_{1}^{+}$$\rightarrow$$0_{1}^{+}$ & 112.8   & 109.4  & 85.2    &  57.4 & 70.0(68)\T\B \\
		&		&	$2_{2}^{+}$$\rightarrow$$0_{1}^{+}$  & 8.1     & 7.1 & 14.2  &  25.4 &  7.8(18)\T\B \\
		&	27	&	$3/2_{1}^{+}$$\rightarrow$$5/2_{1}^{+}$ & 264.3   & 239.7  & 148.5    & 91.4 &  110.7\T\B \\
		& 28 & $2_{1}^{+}$$\rightarrow$$0_{1}^{+}$ & 144.9   & 141.7   & 125.2  &  99.8 & 66.7(25)\T\B \\
		&	29	&	$3/2_{1}^{+}$$\rightarrow$$1/2_{1}^{+}$ &  0.1  & 0.002 & 10.1 & 37.9 & 21.7(21) \T\B \\							
		&	30	&	$2_{1}^{+}$$\rightarrow$$0_{1}^{+}$   & 118.4   & 112.4  & 37.8  &  58.6 &  47.1(61) \T\B \\
		&		&	$1_{1}^{+}$$\rightarrow$$2_{1}^{+}$  & 0.3    & 0.01  & 11.7  & 2.8 &   8.3(61)\T\B \\
		&	31	&	$5/2_{1}^{+}$$\rightarrow$$3/2_{1}^{+}$& 59.0  & 102.2  &   111.8   & 91.6 &   69(23)\T\B \\				
		&	32	&	$2_{1}^{+}$$\rightarrow$$0_{1}^{+}$  & 96.9    & 90.5   & 74.4    &  54.3 &  26.6(78)\T\B \\
		&		&	$2_{2}^{+}$$\rightarrow$$0_{1}^{+}$  & 0.2  & 0.7  & 1.1   &  2.2  &  1.03(36) \T\B \\	
		&	33	&	$1/2_{1}^{+}$$\rightarrow$$3/2_{1}^{+}$  & 137.0    & 123.0& 84.2     &  54.6 &  NA \T\B \\			
		&	34	&	$2_{1}^{+}$$\rightarrow$$0_{1}^{+}$ & 46.9    & 46.1   & 41.6    &  43.2 &  17.0(65) \T\B \\	
		
		P	&	29	& $3/2_{1}^{+}$$\rightarrow$$1/2_{1}^{+}$ & 2.3  & 3.4   & 22.8    & 55.7 &  14.3(26) \T\B \\	
		&	& $5/2_{1}^{+}$$\rightarrow$$1/2_{1}^{+}$ & 83.1  &100.1   & 89.7    & 66.8 &  NA \T\B \\	
		&	30	&	$2_{1}^{+}$$\rightarrow$$1_{1}^{+}$& 31.0     & 39.8   & 60.6     &  7.6 & 0.83(22) \T\B\\
		&	31	 &	 $3/2_{1}^{+}$$\rightarrow$$1/2_{1}^{+}$&  33.0  &  34.4   &  46.1     &   42.9 &   24.3(35) \T\B\\	
		&	&	$5/2_{1}^{+}$$\rightarrow$$1/2_{1}^{+}$& 44.8  & 59.7   & 29.7     &  54.2 &  37.0(29) \T\B\\	
		&	32	& $3_{1}^{+}$$\rightarrow$$1_{1}^{+}$ & 62.8  & 50.8   & 19.1    & 0.3 & NA  \T\B\\
		&	33	& $3/2_{1}^{+}$$\rightarrow$$1/2_{1}^{+}$ & 47.4  & 54.8   & 54.4    &  48.4 & 63(25)  \T\B\\
		&	& $5/2_{1}^{+}$$\rightarrow$$1/2_{1}^{+}$ & 57.5  & 58.8   & 47.1    &  38.4 &  32.1(50) \T\B\\
		&35	& $3/2_{1}^{+}$$\rightarrow$$1/2_{1}^{+}$ & 26.5  & 29.2  & 31.6  &  29.5 & NA  \T\B\\
		&		& $5/2_{1}^{+}$$\rightarrow$$1/2_{1}^{+}$ & 34.1  & 33.0   & 33.2  &  32.2 & NA  \T\B\\	
		\hline
		\hline 
	\end{tabular}
\end{table*}

First, we focus on the B(E2) transition strengths for Si and P isotopes. In Table \ref{B(E2)}, B(E2) transition strengths using microscopic as well as USDB interactions for the selected transitions are reported. Each microscopic interaction predicts enhanced transition strength for  $2_{1}^{+}$$\rightarrow$$0_{1}^{+}$ in the case of \textsuperscript{26,28,30,32,34}Si, except for DJ16A in \textsuperscript{30}Si. Information about collectivity for nuclei can be determined from the B(E2) transition strength. A gradual decrease of B(E2) value for an isotopic chain shows appearance of magic number. Experimentally, as neutrons are increased from $N = 12$ ($^{26}$Si), the B(E2) transition strength from $2^+_1$ to $0^+_1$ decreases in even-even nuclei until the closed shell at $N=20$ is reached. This trend is observed for the JISP16 and N3LO interactions from \textsuperscript{28}Si to \textsuperscript{34}Si.  Although, the B(E2; $2_{1}^{+}$$\rightarrow$$0_{1}^{+}$) results for DJ16A are deviated from the experimental pattern, its values are still closer to the experimental data than other two microscopic interactions. For $^{29}$Si and $^{30}$Si, transitions from $3/2^+_1$ to $1/2^+_1$ and $1^+_1$ to $2^+_1$, respectively, obtained from both JISP16 and N3LO interactions, are significantly weaker than that found experimentally. We have also predicted B(E2) transition strength for $5/2^+_1$$\rightarrow$$1/2^+_1$ ($^{29,35}$P), $3^+_1$$\rightarrow$$1^+_1$ ($^{32}$P) and  $3/2^+_1$$\rightarrow$$1/2^+_1$ (\textsuperscript{33}Si,$^{35}$P). It seems from these data that B(E2) is very sensitive to the details of the wave-functions, thus, further improvement in DJ16A is required. We have also computed the root-mean-square (rms) deviation for B(E2) values and obtained deviations corresponding to N3LO, JISP16, DJ16A, and USDB are 51.55, 47.09, 29.26, and 20.27 $e^2$fm$^4$, respectively. Hence, we can say that the results of B(E2) calculated using DJ16A interaction are closest to the experimental value among the microscopic interactions.

Now, we discuss the electric quadrupole and magnetic dipole  moments of either g.s. or (and) first excited state (additionally, $5/2_1^+$ of $^{31}$P also), which are shown in Table \ref{Moments}. We have selected only those states for which experimental data of either quadrupole or magnetic moment or both are available, to check the predictive power of our theoretical shell model interactions. Calculated quadrupole moment using all interactions for \textsuperscript{28}Si and \textsuperscript{28}P are in reasonable agreement with the experiment. Effective single particle (SP) quadrupole moment of $5/2^+$ corresponding to last unpaired neutron in $d_{5/2}$ orbital is 0.022 eb for $^{27}$Si. Experimental and calculated shell model values are quite larger than the SP value. As discussed earlier that probability of the configuration $\ket{\pi (d_{5/2}^6) \otimes \nu (d_{5/2}^5)}$ is very small, so, collective feature is developed in this state. For \textsuperscript{30}Si, sign of quadrupole moment for $2^+$ is reproduced correctly only by the DJ16A interaction. We also observe a shape change from $^{28}$Si to $^{30}$Si with DJ16A that supports the experimental data. The $^{28}$Si has an oblate deformation, whereas $^{30}$Si is found nearly spherical. For $3/2^+$ in $^{33}$Si, effective SP quadrupole moment is obtained as 0.018 eb. The experimental data is not available to examine the single particle nature of this state. Thus, we have calculated the quadrupole moment with each interaction and noted that the quadrupole moment decreases from N3LO to JISP16 to DJ16A to USDB, still far from the SP value. We have predicted the quadrupole moment for \textsuperscript{31}P and \textsuperscript{32}P that will be helpful in future experimental study. 
\begin{sidewaystable}
	\centering
	\caption{	\label{Moments} Calculated electric quadrupole and magnetic dipole moments of Si and P nuclei.
		Experimental quadrupole moments are taken from \cite{STONE20161} and magnetic moments are from \cite{NNDC,IAEA}. We have used effective charges as $e_p = 1.5e$, $e_n = 0.5e$ and g$_s^\mathrm{eff}$ = g$_s^\mathrm{free}$.	
	}
	\begin{tabular}{cccccccccccccc}
		\vspace{-2.8mm}\\
		\hline \hline
		&                     &     & \multicolumn{5}{c}{Q (eb)}    & \multicolumn{5}{c}{$\mu$ ($\mu_N$)} \T\B\\
		\cline{4-8}
		\cline{9-13}
		Nuclei  & $A$  & J$^{\pi}$ & N3LO& JISP16 & DJ16A & USDB 	& Exp.
		& N3LO & JISP16 & DJ16A & USDB  &	Exp.\T\B\\
		\hline
		
		Si      &		27  & 5/2$^+$    & 0.082     & 0.094     & 0.136      & 0.141 & 0.063(14)
		& 0.477   & 0.298     & -0.261       & -0.678 	& (-)0.8652(4)\T\B\\
		
		&		28 &2$^+$& 0.245  & 0.242    & 0.231  & 0.209 	& 0.16(3)
		& 1.015 & 1.019    & 1.025  & 1.031 & 1.1(2)  \T\B\\
		
		&	29		& 1/2$^+$ & - &- &- &- & -
		& -0.085  & -0.120   & -0.164  & -0.503 & -0.55529(3)\T\B\\
		
		&		30 & 2$^+$	& 0.191 & 0.187  & -0.100    & 0.024 & -0.05(6)
		& 1.141   & 1.187   &  0.392   & 0.732 & 0.8(2) \T\B\\
		
		&		33 & 3/2$^+$  & 0.122 & 0.114  & 0.093    & 0.065 	& NA
		& 1.510   & 1.422   & 1.325    & 1.206 & 1.21(3) \T\B\\
		
		P      &  28	& 3$^+$	& 0.119	& 0.096	& 0.185	 & 0.148 & 0.137(14)
		& 2.180 & 1.192 & -0.497& 0.303 & 0.312(3)  \T\B\\
		&29	& 1/2$^+$	&-&-&-&-&-
		& 0.582 & 0.627 & 0.732 & 1.133 &1.2346(3) \T\B\\
		
		& 31 & 1/2$^+$ & -&- & - & -& -
		& 0.196 &0.395 &0.473  &1.087  & 1.13160(3)\T\B\\   
		&& 3/2$^{+}$   & -0.138    & -0.131   & -0.115  & -0.091 & NA
		& 0.464    & 0.387     & 0.232     &  0.164 &  0.30(8) \T\B\\
		
		&  & 5/2$^+$ & 0.189 &0.197 & 0.193 & 0.049& NA
		& 0.706 &1.010 & 0.987 & 2.217 & 2.8(5)\T\B\\		
		& 32 & 1$^+$ & -0.050& -0.058 & -0.047  & -0.019  & NA
		& -0.657 & -0.382 & -0.239 & -0.021 & -0.2524(3) \T\B\\ 	
		& 35 & 1/2$^+$ & -&- & - & -& -
		& 1.754 &1.829 &1.848  &1.961  & NA\T\B\\ 
		\hline \hline
	\end{tabular}	
\end{sidewaystable}
Magnetic moment of $2^{+}$ using all interactions are in good agreement with the experimental data for $^{28}$Si. The SP magnetic moments for $^{27}$Si, $^{29}$Si and $^{33}$Si are -1.913, -1.913 and 1.148 $\mu_N$ for last neutron in $d_{5/2}$, $s_{1/2}$ and $d_{3/2}$ orbitals, respectively. The sign of magnetic moment for $5/2^+$ state in $^{27}$Si has not been confirmed experimentally yet. The N3LO and JISP16 interactions give positive sign of magnetic moment for g.s., while DJ16A and USDB predict negative sign that is favoured by the sign of SP moment. In case of \textsuperscript{29}Si, our microscopic interactions reproduce the correct sign of magnetic moment for $1/2^+$ state with underestimated values. Since there is large configuration mixing for these states, they show collective behavior. Experimental g.s. magnetic moment of 1.21(3) $\mu_N$ for $^{33}$Si is close to the SP magnetic moment. The $3/2^+$ state has the configuration $\ket{\pi (d_{5/2}^6) \otimes \nu (d_{5/2}^6 s_{1/2}^2 d_{3/2}^3)}$ with 69.14 \% and 77.99 \% of probabilities according to DJ16A and USDB interactions, respectively, so, their magnetic moments of 1.325 and 1.206 $\mu_N$ match well with the SP moment. On the other hand, the magnetic moments for N3LO and JISP16 are slightly deviated from the SP magnetic moment because these interactions generate the same configuration with 38.52 \% and 48.75 \% of probabilities, respectively. $^{33}$Si is one neutron hole nuclei beyond $N=20$ shell closure, thus, it is more likely to show single particle structure. 

Experimental magnetic moment of g.s. for $^{29,31}$P is significantly smaller than the SP magnetic moment (2.793 $\mu_N$), revealing mixed configurations in its wave-functions. Magnetic moment of $1/2^+$ improves from N3LO to JISP16 to DJ16A for $^{29,31}$P, still difference between calculated and experimental value is  large. There is no experimental data available for \textsuperscript{35}P. So, we have given prediction for $1/2^+$ state using all interactions. The g.s. is dominated by the proton configuration of $\ket{\pi (d_{5/2}^6s_{1/2}^1)}$ with 88-92 \% probability using different interactions. SP magnetic moments for $1/2^+$ is 2.793 $\mu_N$ that is not close to the calculated shell model results. Thus, we have used different effective values \cite{PhysRevC.78.064302} as g$^p_l$ = 1.174, g$^p_s$ = 5.00, g$^n_l$ = -0.11, g$^n_s$ = -3.44 and obtained magnetic moment for N3LO, JISP16, DJ16A and USDB are 1.634, 1.696, 1.712 and 1.806 $\mu_N$, respectively. Now, the effective SP moment is 2.5 $\mu_N$ which indicates that difference between the shell model and SP values is reduced. The calculated rms deviation for magnetic moments for Si and P isotopes, which are tabulated in Table \ref{Moments}, are 1.03, 0.78, 0.70, and 0.21 $\mu_N$ for N3LO, JISP16, DJ16A, and USDB, respectively. This indicates that the DJ16A interaction shows minimum deviation from the experimental data.

\section{Spectroscopic factor for $^{24}$S\lowercase{i}} 
\begin{table*}[h]
	\centering
	\caption{\label{24Si_C^2S}Comparison of calculated shell model spectroscopic factor strengths with the experimental data. Experimental excitation energies and C$^2$S of the states of $^{24}$Si are taken from the Ref. \cite{PhysRevLett.122.232701}. Different $l$ and $j$ values for each state of $^{24}$Si are shown in columns III and IV.}
	\begin{tabular}{ccccccccc}
		\vspace{-2.8mm}\\
		\hline \hline
		\vspace{-2.8mm}\\
		$E_x$ & $J^{\pi}$ & $l$ & $j$ & \multicolumn{5}{c}{C$^2$S} \\
		\cline{5-9}
		\vspace{-2.8mm}\\
		KeV& &  & & N3LO& JISP16  & DJ16A & USDB & Exp.\\
		\hline \vspace{-2.8mm}\\
		0 & 0$_1^+$ & 2 & 5/2 & 2.49& 2.32 & 2.92& 3.42  & $\leq$ 2.8    \\ 
		1874(3) & 2$_1^+$ & 0 & 1/2 & 0.07& 0.26     & 0.28 & 0.25  & 0.6(2)    \\
		&         & 2 & 3/2 & 0.01& 0.02   & 0.009 & 0.03  & 0.07(2)     \\
		&         & 2 & 5/2 & 0.0002& 0.04  & 0.16    & 0.18  & 0.4(1)\\
		3449(5) &(2$_2^+$)& 0 & 1/2 & 0.45 & 0.36   & 0.36  & 0.46 & 0.7(4)     \\
		&         & 2 & 3/2 & 0.0007& 0.00 & 0.03   & 0.00  & 0.002(1)  \\
		&         & 2 & 5/2 & 0.13& 0.16    & 0.27 & 0.16   & 0.3(2)      \\  
		3471(6) &(4$_1^+$)& 2 & 3/2 & 0.009& 0.0001   & 0.03  & 0.02 & 0.07(4)   \\	
		&         & 2 & 5/2 & 0.16& 0.08  & 0.00 & 0.001  & 0.004(3)     \\
		&(0$_2^+$)& 2 & 5/2 & 0.23& 0.47   & 0.17 & 0.24   & 0.8(4)       \\ 
		\hline \hline	  	
	\end{tabular}
\end{table*}

In the present study, spectroscopic factor strengths for g.s. and excited states of $^{24}$Si from one proton transfer reaction $^{23}$Al($d,n$)$^{24}$Si are calculated within the shell model using microscopic effective N3LO, JISP16, DJ16A and empirical USDB interactions. These results are compared with the extracted experimental values \cite{PhysRevLett.122.232701} in table~\ref{24Si_C^2S}. One proton is captured in $l$ = 0 and 2 orbitals to populate the states of $^{24}$Si. Some states of $^{24}$Si are generated by capturing one-proton in $d_{5/2}$ orbital only and some, from the mixing of  $d_{5/2}$ and (or) $s_{1/2}$ and (or) $d_{3/2}$ orbitals.
In Ref. \cite{PhysRevLett.122.232701}, experimental spectroscopic factors (${\rm C^{2}S}$), which are shown in last column of Table \ref{24Si_C^2S}, have been extracted using the relation:
\begin{equation}
	{\rm C^{2}S}_{\rm{exp}}^i = \frac{{\rm C^{2}S}_{\rm{theo}}^i \times \sigma^i_{\rm{theo}}}{\sum_{i'}({\rm C^{2}S}_{\rm{theo}}^{i'} \times \sigma^{i'}_{\rm{theo}})} \times \frac{\sigma_{\rm{exp}}}{\sigma^{i}_{\rm{theo}}},
\end{equation}

where, C$^{2}$S$_{\rm{theo}}$ were calculated corresponding to USDB-cdpn interaction using the shell model \cite{PhysRevC.74.034315} and $i$ represents individual quantum numbers. Extracted experimental C$^{2}$S \cite{PhysRevLett.122.232701} for 0$_1^+$ state is $\leq$2.8 which is in good agreement with previous experimental value 2.7(2) of C$^{2}$S \cite{PhysRevC.86.015806}. The N3LO, JISP16 and DJ16A interactions give the C$^{2}$S as 2.49, 2.32 and 2.92, respectively, while the C$^{2}$S  3.42 is obtained with USDB which is quite away from the experimental value. The calculated C$^{2}$S for g.s. of $^{24}$Si is very large which shows single-particle nature of this state. Experimental excited states of $^{24}$Si are tentative.   For first $2^+$, experimentally extracted C$^{2}$S is much smaller for $d_{3/2}$ transfer than $d_{5/2}$ and $s_{1/2}$. Except for N3LO interaction, all interactions predict smallest value for $d_{3/2}$ transfer. The calculated C$^{2}$S($s_{1/2}$) for $(2_2^+)$ state is in reasonable agreement with experimental value. For $(0_2^+)$ state, theoretical C$^{2}$S is slightly smaller than experimental data.

In the recent work \cite{PhysRevC.101.031303}, spin-parity assignment has been made for excited states of $^{24}$Si. The second excited state of $^{24}$Si was confirmed to be $2_2^+$. Also, the state at 3471 keV of $^{24}$Si is confirmed as $0_2^+$ instead of $4_1^+$.

\section{Conclusion}
In the present work, we have implemented the nuclear shell model with newly constructed microscopic effective interactions to study Si and P isotopes with neutron number $N = 9 - 20$. These microscopic NN interactions, which are obtained from the NCSM wave functions and OLS transformation method, are chiral N3LO, JISP16 and DJ16A. Microscopic results are compared with phenomenological results using USDB interaction and experimentally measured data. We have studied the energy spectra of Si and P isotopes lying in $sd$-shell. We found that monopole modified DJ16 interaction (DJ16A) results for g.s. energies are better than the other microscopic interactions and the proton subshell closure at $Z=14$ persists in Si. Based on the rms deviations, we found that the minimum rms deviation is obtained for DJ16A interaction for both B(E2) and magnetic moment among all microscopic interactions. So, we can say that the DJ16A interaction is a more suitable interaction for these Si and P isotopes. Still, some corrections are needed in DJ16A to reproduce the experimental data. Our calculated results for upper $sd$ shell nuclei will be helpful to further investigate the modifications such as T = 0 monopole components, quadrupole-quadrupole and pairing correlations that should be incorporated in this microscopic interaction.  Also, the deficiencies of N3LO and JISP16 interactions in describing the Si and P isotopes are due to absence of 3N interaction and tuned up to A=16 nuclei only, respectively, which indicate the need of the further tuning of the matrix elements at least to the light $sd$-shell nuclei.

Variation of proton effective single-particle energies for Si isotopes are presented using all the interactions to determine the shell evolution. To understand the contribution from different components of the TBMEs in the shell evolution, we have also studied the spin-tensor decomposition of the effective interactions. Apart from low-lying energy spectra, electromagnetic observables have  been also calculated for complete description of these nuclei. Spectroscopic factor strengths for $^{24}$Si are calculated, which is of importance in astrophysical scenario. It is reported that the C$^{2}$S for g.s. with microscopic interactions is better reproduced than that obtained from USDB interaction.

\section*{ACKNOWLEDGMENTS}
We acknowledge a research grant from SERB (India),  CRG/2019/000556.  P.C. acknowledges the financial support from the MHRD (Government of India) for her Ph.D. thesis work. 
We would like to thank N. A. Smirnova for her valuable suggestions.
We acknowledge National Supercomputing Mission (NSM) for providing computing resources of `PARAM Ganga' at Indian Institute of Technology Roorkee, which is implemented by C-DAC and supported by the Ministry of Electronics and Information Technology (MeitY) and Department of Science and Technology (DST), Government of India.

\bibliography{utphys}
\bibliography{references}

\begin{thebibliography}{70}
\expandafter\ifx\csname natexlab\endcsname\relax\def\natexlab#1{#1}\fi
\expandafter\ifx\csname bibnamefont\endcsname\relax
  \def\bibnamefont#1{#1}\fi
\expandafter\ifx\csname bibfnamefont\endcsname\relax
  \def\bibfnamefont#1{#1}\fi
\expandafter\ifx\csname citenamefont\endcsname\relax
  \def\citenamefont#1{#1}\fi
\expandafter\ifx\csname url\endcsname\relax
  \def\url#1{\texttt{#1}}\fi
\expandafter\ifx\csname urlprefix\endcsname\relax\def\urlprefix{URL }\fi
\providecommand{\bibinfo}[2]{#2}
\providecommand{\eprint}[2][]{\url{#2}}

\bibitem{BARRETT2013131} B. R. Barrett, P. Navr\'atil and J. P. Vary, \textit{Ab initio} no-core shell model, 
\href{https://doi.org/10.1016/j.ppnp.2012.10.003}
{Prog. Part. Nucl. Phys. {\bf 69}, 131 (2013)}.

\bibitem{PhysRevC502841}D. C. Zheng, J. P. Vary and B. R. Barrett, Large-space shell model calculations for light nuclei,
\href{https://doi.org/10.1103/physrevc.50.2841}
{Phys. Rev. C {\bf 50}, 2841 (1994)}.

\bibitem{PhysRevC542986}P. Navr\'atil and B. R. Barrett, No-core shell model calculations with starting-energy-independent multivalued effective interactions, 
\href{https://doi.org/10.1103/PhysRevC.54.2986}
{Phys. Rev. C {\bf 54}, 2986 (1996)}.

\bibitem{PhysRevC573119}P. Navr\'atil and B. R. Barrett, Large-basis shell model calculations for $p$-shell nuclei,
\href{https://doi.org/10.1103/PhysRevC.57.3119}
{Phys. Rev. C {\bf 57}, 3119 (1998)}.

\bibitem{PRC1906} P. Navr\'atil and B. R. Barrett,
Four-nucleon shell model calculations in a Faddeev-like approach,
\href{https://doi.org/10.1103/PhysRevC.59.1906}
{ Phys. Rev. C {\bf 59}, 1906 (1999)}.

\bibitem{PRC62} P. Navr\'atil, J. P. Vary and B. R. Barrett, Large-basis \textit{ab initio} no-core shell model and its application to $^{12}$C,
\href{https://doi.org/10.1103/PhysRevC.62.054311}
{Phys. Rev. C {\bf 62}, 054311 (2000)}.

\bibitem{2009}P. Navr\'atil, S. Quaglioni, I. Stetcu and B. R. Barrett, 
Recent developments in no-core shell model calculations,
\href{https://doi.org/10.1088/0954-3899/36/8/083101}
{ J. Phys. G: Nucl. Part. Phys. {\bf 36}, 083101 (2009)}.

\bibitem{MVS2009} P. Maris, J. P. Vary, and A. M. Shirokov, \textit{Ab initio} no-core full configuration calculations of light nuclei,
\href{https://journals.aps.org/prc/abstract/10.1103/PhysRevC.79.014308}
{Phys. Rev. C {\bf 79}, 014308 (2019)}.

\bibitem{stetcu1}
I. Stetcu, B. R. Barrett, P. Navr\'atil, and J. P. Vary,
Long- and short-range correlations in the ab-initio no-core shell model,
\href{https://doi.org/10.1103/PhysRevC.73.037307}
{Phys. Rev. C 73, 037307 (2006)}.

\bibitem{stetcu2}
I. Stetcu, B. R. Barrett, P. Navr\'atil, and J. P. Vary,
Effective operators within the \textit{ab initio} no-core shell model,
\href {https://link.aps.org/doi/10.1103/PhysRevC.71.044325}
{Phys. Rev. C 71, 044325 (2005)}.

\bibitem{PhysRevC.102.044309} P. Choudhary, P. C. Srivastava and P. Navr\'atil, \textit{Ab initio} no-core shell model study of $^{10-14}\mathrm{B}$  isotopes with realistic NN interactions, 
\href{https://doi.org/10.1103/PhysRevC.102.044309}
{Phys. Rev. C  {\bf 102}, 044309 (2020)}.


\bibitem{PhysRevC.107.014309} P. Choudhary, P. C. Srivastava, M. Gennari and P. Navr\'{a}til, \textit{Ab initio} no-core shell model description of $^{10-14}$C isotopes, 
\href{https://link.aps.org/doi/10.1103/PhysRevC.107.014309}
{Phys. Rev. C {\bf 107}, 014309 (2023)}.

\bibitem{Nucl.Phys.A} P. Choudhary and P. C. Srivastava, \textit{Ab initio} no-core shell model study of neutron-rich $^{18,19,20}$C isotopes,
\href{https://doi.org/10.1016/j.nuclphysa.2022.122565}
{Nucl. Phys. A {\bf 1029}, 122565 (2023)}.

\bibitem{JPG} C. Sarma and P. C. Srivastava, \textit{Ab initio} no-core shell model study of $^{18-24}$Ne isotopes,
\href{https://iopscience.iop.org/article/10.1088/1361-6471/acb962}
{J Phys. G: Nucl. Part. Phys. {\bf 50}, in press (2023)}.

\bibitem{10.1093/ptep/ptz073} A. Saxena and P. C. Srivastava, \textit{Ab initio} no-core shell model study of neutron-rich nitrogen isotopes, 
\href{https://doi.org/10.1093/ptep/ptz073}
{Prog. Theor. Exp. Phys. {\bf 2019}, 073D02 (2019)}.

\bibitem{Saxena_2020} A. Saxena and P. C. Srivastava, \textit{Ab initio} no-core shell model study of $^{18-23}$O and $^{18-24}$F isotopes', 
\href{https://doi.org/10.1088/1361-6471/ab6f1d}
{J Phys. G: Nucl. Part. Phys. {\bf 47}, 055113 (2020)}.



\bibitem{PhysRevLett.113.142501}
S. K. Bogner, H. Hergert, J. D. Holt, A. Schwenk, S. Binder, A. Calci, J. Langhammer, and R. Roth, Nonperturbative shell model interactions from the in-medium similarity renormalization group,
\href{https://journals.aps.org/prl/abstract/10.1103/PhysRevLett.113.142501}{ Phys. Rev. Lett. {\bf 113}, 142501 (2014)}.

\bibitem{PhysRevLett.106.222502}
K. Tsukiyama, S. K. Bogner, and A. Schwenk, In-medium similarity renormalization group for nuclei,
\href{https://journals.aps.org/prl/abstract/10.1103/PhysRevLett.106.222502}{Phys. Rev. Lett. {\bf 106}, 222502 (2011)}.

\bibitem{PhysRevLett.113.142502}
G. R. Jansen, J. Engel, G. Hagen, P. Navr\'atil, and A. Signoracci, \textit{Ab initio} coupled-cluster effective interactions for the shell model: Application to neutron-rich oxygen and carbon isotopes,
\href{https://journals.aps.org/prl/abstract/10.1103/PhysRevLett.113.142502}{Phys. Rev. Lett. {\bf 113}, 142502 (2014)}.

\bibitem{PhysRevC.94.011301}
G. R. Jansen, M. D. Schuster, A. Signoracci, G. Hagen, and P. Navr\'atil, Open $sd$-shell nuclei from first principles,
\href{https://journals.aps.org/prc/abstract/10.1103/PhysRevC.94.011301}{Phys. Rev. C {\bf 94}, 011301(R) (2016)}.

\bibitem{Launey}
K. D. Launey, T. Dytrych, and J. P. Draayer, Symmetry-guided large-scale shell model theory, 
\href{https://doi.org/10.1016/j.ppnp.2016.02.001}
{Prog. Part. Nucl. Phys. {\bf 89}, 101 (2016)}.

\bibitem{Panu1}
P. Ruotsalainen, J. Henderson, G. Hackman, G. H. Sargsyan,
K. D. Launey, A. Saxena, P. C. Srivastava, S. R. Stroberg,
T. Grahn, J. Pakarinen, \textit{et al.},
Isospin symmetry in $B(E2)$ values : Coulomb excitation study of $^{21}$Mg,
\href{https://doi.org/10.1103/PhysRevC.99.051301}
{Phys. Rev. C {\bf 99}, 051301(R) (2019)}.

\bibitem{Jonathan}
 J. Williams, G. C. Ball, A. Chester, T. Domingo, A. B.
Garnsworthy, G. Hackman, J. Henderson, R. Henderson, R.
Krücken, A. Kumar, K. D. Launey, J. Measures, O. Paetkau,
J. Park, G. H. Sargsyan, J. Smallcombe, P. C. Srivastava, K.
Starosta, \textit{et al.}, Structure of 
$^{28}$Mg and influence of the neutron $pf$-shell, 
\href{https://doi.org/10.1103/PhysRevC.100.014322}
{Phys. Rev. C {\bf 100}, 014322 (2019)}.

\bibitem{SA-NCSM}
O. M. Molchanov, K. D. Launey, A. Mercenne, G. H. Sargsyan, T. Dytrych, and J. P. Draayer, 
Machine learning approach to pattern recognition in nuclear dynamics
from the \textit{ab initio} symmetry-adapted no-core shell model,
\href{https://journals.aps.org/prc/pdf/10.1103/PhysRevC.105.034306}
{Phys. Rev. C {\bf 105}, 034306 (2022)}.

\bibitem{PhysRevC.96.024316}
A. Saxena and P. C. Srivastava, First-principles results for electromagnetic properties of $sd$-shell nuclei,
\href{{https://link.aps.org/doi/10.1103/PhysRevC.96.024316}}
{Phys. Rev. C {\bf 96}, 024316 (2017)}.

\bibitem{Collectivity}
A. Saxena, A. Kumar, V. Kumar,P. C. Srivastava and T. Suzuki,  \textit{Ab initio} description of collectivity for $sd$-shell nuclei,
\href{https://doi.org/10.1007/s10751-019-1582-y}
{Hyperfine Interact. {\bf 240}, 37 (2019)}. 

\bibitem{PhysRevC.105.034333}
S. R. Stroberg, J. Henderson, G. Hackman, P. Ruotsalainen, G. Hagen, and J. D. Holt, Systematics of $E2$ strength in the $sd$-shell with the valence-space in-medium similarity renormalization group,
\href{10.1103/PhysRevC.105.034333}
{Phys. Rev. C {\bf 105}, 034333 (2022)}.

\bibitem{PhysRevC.105.034332}
J. Henderson, G. Hackman, P. Ruotsalainen, J. D. Holt, S. R. Stroberg, C. Andreoiu, G. C. Ball, N. Bernier, M. Bowry, R. Caballero-Folch,  \textit{et al.}, Coulomb excitation of the $|{T}_{z}|=\frac{1}{2}, A=23$ mirror pair,
\href{10.1103/PhysRevC.105.034332}
{Phys. Rev. C {\bf 105}, 034332 (2022)}.

\bibitem{SF}
P. C. Srivastava and V. Kumar,
Spectroscopic factor strengths using \textit{ab initio} approaches,
\href{https://journals.aps.org/prc/pdf/10.1103/PhysRevC.94.064306}
{Phys. Rev. C {\bf 94}, 064306 (2016)}.

\bibitem{GT}
A. Saxena, P. C. Srivastava, and T. Suzuki, \textit{Ab initio} calculations of Gamow-Teller strengths in the $sd$-shell,
\href{10.1103/PhysRevC.97.024310}
{Phys. Rev. C {\bf 97}, 024310 (2018)}. 

\bibitem{PhysRevC.78.044302}
A. F. Lisetskiy, B. R. Barrett, M. K. G. Kruse, P. Navr\'atil, I. Stetcu, and J. P. Vary, \textit{Ab-initio} shell model with a core, 
\href{https://journals.aps.org/prc/abstract/10.1103/PhysRevC.78.044302}{Phys. Rev. C {\bf 78}, 044302 (2008)}.

\bibitem{PhysRevC.91.064301}
E. Dikmen, A. F. Lisetskiy, B. R. Barrett, P. Maris, A. M. Shirokov, and J. P. Vary, \textit{Ab initio} effective interactions for $sd$-shell valence nucleons,
\href{https://journals.aps.org/prc/abstract/10.1103/PhysRevC.91.064301}{Phys. Rev. C {\bf 91}, 064301 (2015)}.

\bibitem{PhysRevC.100.054329}
N. A. Smirnova, B. R. Barrett, Y. Kim, I. J. Shin, A. M. Shirokov, E. Dikmen, P. Maris, and J. P. Vary, Effective interactions in the $sd$-shell,
\href{https://journals.aps.org/prc/abstract/10.1103/PhysRevC.100.054329}
{Phys. Rev. C {\bf 100}, 054329 (2019)}.

\bibitem{SHIROKOV200733}
A. M. Shirokov and J.P. Vary and A.I. Mazur and T.A. Weber,
Realistic nuclear Hamiltonian: Ab exitu approach,
\href{https://doi.org/10.1016/j.physletb.2006.10.066}{Phys. Lett. B {\bf 644}, 33 (2007)}.

\bibitem{PhysRevC.68.041001} D. R. Entem and R. Machleidt, Accurate charge-dependent nucleon-nucleon potential at fourth order of chiral perturbation theory, 
\href{https://doi.org/10.1103/PhysRevC.68.041001}
{Phys. Rev. C {\bf 68}, 041001(R) (2003)}.

\bibitem{A.M.Shirokov} A. M. Shirokov, I. J. Shin, Y. Kim, M. Sosonkina, P. Maris, and J. P. Vary, N$^3$LO NN interaction adjusted to light nuclei in \textit{ab exitu} approach,
\href{https://doi.org/10.1016/j.physletb.2016.08.006} 
{Phys. Lett. B {\bf 761}, 87 (2016)}.

\bibitem{P.D.Cottle}
P. D. Cottle, Single proton energies in the Si isotopes and the $Z=14$ subshell closure, 
\href{10.1103/PhysRevC.76.027301}
{Phys. Rev. C {\bf 76}, 027301 (2007)}.

\bibitem{SPEs}
O. V. Bespalova,N. A. Fedorov, A. A. Klimochkina, M. L. Markova, T. I. Spasskaya and T. Yu. Tretyakova, Evolution of single-particle structure of silicon isotopes, 
\href{https://doi.org/10.1140/epja/i2018-12449-x}{Eur. Phys. J. A {\bf 54}, 2 (2018)}. 

\bibitem{Kaneko}
K. Kaneko, S. Yang, T. Mizusaki, and M. Hasegawa, Shell-model study for neutron-rich $sd$-shell nuclei,
\href{10.1103/PhysRevC.83.014320}{Phy. Rev. C {\bf 83}, 014320 (2011)}.

\bibitem{Kumar:2021rsv}
P. Kumar, V. Thakur, S. Thakur, V. Kumar, and S. K. Dhiman, Evolution of Nuclear Shapes in Light Nuclei from Proton- to Neutron-rich Side,
\href{https://www.actaphys.uj.edu.pl/R/52/5/401/pdf}{Acta Phys. Pol. B {\bf 52}, 401 (2021)}.

\bibitem{PhysRevC.74.034315}
B. A. Brown and W. A. Richter, New ``usd'' hamiltonians for the $\mathit{sd}$ shell, \href{https://journals.aps.org/prc/abstract/10.1103/PhysRevC.74.034315}{Phys. Rev. C {\bf 74}, 034315 (2006)}.

\bibitem{PhysRevC.78.064302}
W. A. Richter, S. Mkhize, and B. A. Brown, $\mathit{sd}$-shell observables for the USDA and USDB hamiltonians,
\href{https://doi.org/10.1103/PhysRevC.78.064302}
{Phys. Rev. C {\bf 78}, 064302 (2008)}.

\bibitem{PhysRevC.101.064312}
A. Magilligan and B. A. Brown, New isospin-breaking ``usd'' hamiltonians for the sd shell, \href{https://doi.org/10.1103/PhysRevC.101.064312}
{ Phys. Rev. C {\bf 101}, 064312 (2020)}.

\bibitem{PhysRevC.95.021304}
N. Tsunoda, T. Otsuka, N. Shimizu, M. Hjorth-Jensen, K. Takayanagi, and T. Suzuki, Exotic neutron-rich medium-mass nuclei with realistic nuclear forces,
\href{https://doi.org/10.1103/PhysRevC.95.021304}{Phys. Rev. C {\bf 95}, 021304 (2017)}.

\bibitem{RMP}E. Epelbaum, H.-W. Hammer and U.-G. Mei\ss ner, Modern theory of nuclear forces, 
\href{https://doi.org/10.1103/RevModPhys.81.1773}
{Rev. Mod. Phys. {\bf 81}, 1773 (2009)}.

\bibitem{EFT1}S. Weinberg, Phenomenological lagrangians, 
\href{https://doi.org/10.1016/0378-4371(79)90223-1}
{Physica A {\bf 96}, 327 (1979)}.

\bibitem{EFT2}S. Weinberg, Nuclear forces from chiral lagrangians, 
\href{https://doi.org/10.1016/0370-2693(90)90938-3}
{Phys. Lett. B {\bf 251}, 288 (1990)}.

\bibitem{EFT3}S. Weinberg, Effective chiral lagrangians for nucleon-pion interactions and nuclear forces, 
\href{https://doi.org/10.1016/0550-3213(91)90231-L}
{ Nucl. Phys. B {\bf 363}, 3 (1991)}.

\bibitem{QCD} R. Machleidt and D. R. Entem, Chiral effective field theory and nuclear forces,  
\href{https://doi.org/10.1016/j.physrep.2011.02.001}
{Phys. Rep. {\bf 503}, 1 (2011)}.

\bibitem{TAKAYANAGI201191}
K. Takayanagi, Effective hamiltonian in the extended krenciglowa–kuo method,
\href{https://doi.org/10.1016/j.nuclphysa.2011.06.025}{Nucl. Phys. A {\bf 864}, 91-112 (2011)}

\bibitem{TAKAYANAGI201161}
K. Takayanagi, Effective interaction in non-degenerate model space,
\href{https://doi.org/10.1016/j.nuclphysa.2011.01.003}{Nucl. Phys. A {\bf 852}, 61--81 (2011)}.

\bibitem{PhysRevC.89.024313}
N. Tsunoda, K. Takayanagi, M. Hjorth-Jensen, and T. Otsuka, Multi-shell effective interactions, \href{https://doi.org/10.1103/PhysRevC.89.024313}{Phys. Rev. C {\bf 89}, 024313 (2014)}.

\bibitem{10.1143/PTP.17.360}
J. Fujita and H. Miyazawa, Pion Theory of Three-Body Forces,
\href{https://doi.org/10.1143/PTP.17.360}{Prog. Theor. Phys. {\bf 17}, 360-365  (1957)}.

\bibitem{Tsunoda:2020gpt}
N. Tsunoda, T. Otsuka, K. Takayanagi, N. Shimizu, T. Suzuki, Y. Utsuno, S. Yoshida, and H. Ueno, The impact of nuclear shape on the emergence of the neutron dripline, \href{https://doi.org/10.1038/s41586-020-2848-x}{Nature {\bf 587}, 66--71 (2020)}.

\bibitem{PhysRevC.105.014319}
T. Otsuka, N. Shimizu, and Y. Tsunoda, Moments and radii of exotic Na and Mg isotopes, \href{https://doi.org/10.1103/PhysRevC.105.014319}
{Phys. Rev. C {\bf 105}, 014319 (2022)}.

\bibitem{PhysRevC.86.015806}
A. Banu, F. Carstoiu, N. L. Achouri, W. N. Catford, M. Chartier,
  B. Fern\'andez-Dom\'{\i}nguez, M. Horoi, B. Laurent, N. A. Orr, S. Paschalis \textit{et al.}, One-proton breakup of ${}^{24}$Si and the
  ${}^{23}$Al($p,\ensuremath{\gamma}$)${}^{24}$Si reaction in type I x-ray
  bursts,
  \href{https://journals.aps.org/prc/abstract/10.1103/PhysRevC.86.015806}{ Phys. Rev. C {\bf 86}, 015806 (2012)}.

\bibitem{PhysRevLett.122.232701}
C. Wolf, C. Langer, F. Montes, J. Pereira, W.-J. Ong, T. Poxon-Pearson, S. Ahn,
S. Ayoub, T. Baumann, D. Bazin \textit{et al.}, Constraining the neutron star compactness: Extraction of the $^{23}\mathrm{Al}\mathbf{(}p,\ensuremath{\gamma}\mathbf{)}$ reaction rate for
the $rp$ process,
\href{https://journals.aps.org/prl/abstract/10.1103/PhysRevLett.122.232701}{Phys. Rev. Lett. {\bf 122}, 232701 (2019)}.

\bibitem{PhysRevC.63.024001}
R. Machleidt, High-precision, charge-dependent bonn nucleon-nucleon potential,
\href{https://doi.org/10.1103/PhysRevC.63.024001}
{Phys. Rev. C {\bf 63}, 024001 (2001)}.

\bibitem{Prog.Theor.Phys.12} S. \^{O}kubo,  Diagonalization of Hamiltonian and Tamm-Dancoff equation, 
\href{https://doi.org/10.1143/PTP.12.603}
{ Prog. Theor. Phys. {\bf 12}, 603 (1954)}.

\bibitem{Prog.Theor.Phys.} K. Suzuki and S. Y. Lee, Convergent theory for effective interaction in nuclei, 
\href{https://doi.org/10.1143/PTP.64.2091}
{Prog. Theor. Phys. {\bf 64}, 2091 (1980)}.

\bibitem{Prog.Theor.Phys.68} K. Suzuki, Construction of Hermitian effective interaction in nuclei: General relation between Hermitian and non-Hermitian forms, 
\href{https://doi.org/10.1143/PTP.68.246}
{Prog. Theor. Phys. {\bf 68}, 246 (1982)}.

\bibitem{Prog.Theor.Phys.92} K. Suzuki and R. Okamoto, Effective interaction theory and unitary-model-operator approach to nuclear saturation problem,
\href{https://doi.org/10.1143/ptp/92.6.1045}
{Prog. Theor. Phys. {\bf 92}, 1045 (1994)}.

\bibitem{PhysRevC.75.061001}
S. K. Bogner, R. J. Furnstahl, and R. J. Perry, Similarity renormalization group for nucleon-nucleon interactions,
\href{10.1103/PhysRevC.75.061001}{Phys. Rev. C {\bf 75}, 061001 (2007)}.

\bibitem{KSHELL2019}N. Shimizu, T. Mizusaki, Y. Utsuno and Y. Tsunoda, Thick-restart block Lanczos method for large-scale shell-model calculations, 
\href{https://doi.org/10.1016/j.cpc.2019.06.011}
{Comput. Phys. Comm. {\bf 244}, 372 (2019)}.
	
\bibitem{NNDC}Data extracted using the National nuclear data center World Wide Web site from the evaluated nuclear structure data file, 
\href{https://www.nndc.bnl.gov/ensdf/.}
{ https://www.nndc.bnl.gov/ensdf/}.	

\bibitem{BaFr64}
R.~K. Bansal and J.~B. French, Even-parity-hole states in $f_{7/2}$-shell nuclei,
\href{https://doi.org/10.1016/0031-9163(64)90648-1}{Phys. Lett. {\bf 11},  145  (1964)}.

\bibitem{PoZu81}
A. Poves and A. Zuker, Theoretical spectroscopy and the fp shell,
\href{https://www.sciencedirect.com/science/article/pii/0370157381901538}{Phys. Rep. {\bf 70}, 235 (1981)}.

\bibitem{Ot01}
T. Otsuka, M. Honma, T. Mizusaki, N. Shimizu and Y. Utsuno, Monte Carlo shell model for atomic nuclei,
\href{https://www.sciencedirect.com/science/article/pii/S0146641001001570}{Prog. Part. Nucl. Phys. {\bf 47}, 319 (2001)}.

\bibitem{UmMu04}
A. Umeya and K. Muto, Triplet-even channel attraction for shell gaps, 
\href{https://link.aps.org/doi/10.1103/PhysRevC.69.024306}{Phys. Rev. C {\bf 69},  024306 (2004)}.

\bibitem{monopole1}
N. A. Smirnova, B. Bally, K. Heyde, F. Nowacki and K. Sieja, Shell evolution and nuclear forces,
\href{https://biblio.ugent.be/publication/1111685/file/1112053.pdf}
{Phys. Lett. B {\bf 686}, 109 (2010).}

\bibitem{monopole}
T. Otsuka, A. Gade, O. Sorlin, T. Suzuki and Y. Utsuno,
Evolution of shell structure in exotic nuclei,
\href{10.1103/RevModPhys.92.015002}
{Rev. Mod. Phys. {\bf 92}, 015002 (2020)}.

\bibitem{monopole2}
T. Otsuka and Y. Tsunoda, The role of shell evolution in shape coexistence, 
\href{https://iopscience.iop.org/article/10.1088/0954-3899/43/2/024009}
{J. Phys. G: Nucl. Part. Phys. {\bf 43}, 024009 (2016)}.

\bibitem{monopole3}
T. Otsuka, R. Fujimoto, Y. Utsuno, B. A. Brown, M. Honma, and T. Mizusaki, Magic Numbers in Exotic Nuclei and Spin-Isospin Properties of the NN Interaction,
\href{10.1103/PhysRevLett.87.082502}
{Phys. Rev. Lett. {\bf 87}, 082502 (2001)}.

\bibitem{monopole4}
T. Otsuka, T. Suzuki, R. Fujimoto, H. Grawe and Y. Akaishi, Evolution of Nuclear Shells due to the Tensor Force,
\href{https://journals.aps.org/prl/pdf/10.1103/PhysRevLett.95.232502}
{Phys. Rev. Lett. {\bf 95}, 232502 (2005)}.

\bibitem{monopole5}
T. Otsuka, M. Honma, T. Mizusaki, N. Shimizu, Y. Utsuno,
Monte Carlo shell model for atomic nuclei,
\href{https://doi.org/10.1016/S0146-6410(01)00157-0}
{Prog. Part. Nucl. Phys. {\bf 47}, 319 (2001)}.

\bibitem{monopole6}
P.C. Srivastava and I. Mehrotra,
Monopole shift in odd neutron-rich F isotopes: A shell model description,
\href{https://doi.org/10.1142/S0218301311018125}
{Int. J. Mod. Phys. E {\bf 20}, 637 (2011)}.

\bibitem{Kir73}
M.~W. Kirson, Spin-tensor decomposition of nuclear effective interactions,
\href{https://doi.org/10.1016/0370-2693(73)90582-0}{Phys. Lett. B {\bf 51},  110  (1973)}.

\bibitem{KlKn77}
K. Klingenbeck, W. Kn\"upfer, M. G. Huber, and P. W. M. Glaudemans, Central and noncentral components of the effective $\mathrm{sd}$-shell interaction, 
\href{https://link.aps.org/doi/10.1103/PhysRevC.15.1483}{Phys. Rev. C {\bf 15},  1483  (1977)}.

\bibitem{STONE20161}
N.J. Stone, Table of nuclear electric quadrupole moments,
\href{https://doi.org/10.1016/j.adt.2015.12.002}
{Atomic Data and Nuclear Data Tables {\bf 111}, 1 (2016)}.

\bibitem{IAEA}IAEA, 
\href{https://www-nds.iaea.org/nuclearmoments/.}
{https://www-nds.iaea.org/nuclearmoments/}.

\bibitem{PhysRevC.101.031303}
B. Longfellow, A. Gade, J. A. Tostevin, E. C. Simpson, B. A. Brown, A. Magilligan, D. Bazin, P. C. Bender, M. Bowry, B. Elman \textit{et al.}, Two-neutron knockout as a probe of the composition of states in $^{22}\mathrm{Mg},^{23}\mathrm{Al}$, and $^{24}\mathrm{Si}$, \href{https://doi.org/10.1103/PhysRevC.101.031303}{Phys. Rev. C {\bf 101}, 031303 (2020)}.


\end{thebibliography}

\end{document}